\documentstyle[12pt,aasms4]{article}
\slugcomment{Accepted  to be published in The Astrophysical Journal}

\lefthead{Gonz\'alez Delgado et al.}
\righthead{Seyferts 2}

\begin{document}

\title{Ultraviolet-Optical observations of the Seyfert 2 Galaxies 
NGC 7130, NGC 5135 and IC 3639: Implications for the Starburst-AGN Connection.}

\author{Rosa M. Gonz\'alez Delgado}
\affil{Space Telescope Science Institute, 3700 San Martin Drive, Baltimore, MD
21218}
\affil{Instituto de Astrof\'\i sica de Andaluc\'\i a, Apdo. 3004, 18080 Granada, Spain}
\affil{Electronic mail: rosa@iaa.es, gonzalez@stsci.edu}

\author{Timothy Heckman}
\affil{Department of Physics \& Astronomy, JHU, Baltimore, MD 21218}
\affil{Adjunct Astronomer at STScI}
\affil{Electronic mail: heckman@pha.jhu.edu}

\author{Claus Leitherer}
\affil{Space Telescope Science Institute, 3700 San Martin Drive, Baltimore, MD
21218}
\affil{Electronic mail: leitherer@stsci.edu}

\author{Gerhardt Meurer}
\affil{Department of Physics \& Astronomy, JHU, Baltimore, MD 21218}
\affil{Electronic mail: meurer@pha.jhu.edu}

\author{Julian Krolik}
\affil{Department of Physics \& Astronomy, JHU, Baltimore, MD 21218}
\affil{Electronic mail: jhk@pha.jhu.edu}

\author{Andrew S. Wilson}
\affil{Astronomy Department, University of Maryland, College Park, MD 20742}
\affil{Adjunct Astronomer at STScI}
\affil{Electronic mail: wilson@astro.umd.edu}

\author{Anne Kinney}
\affil{Space Telescope Science Institute, 3700 San Martin Drive, Baltimore, MD
21218}
\affil{Electronic mail: kinney@stsci.edu}

\and

\author{Anuradha Koratkar}
\affil{Space Telescope Science Institute, 3700 San Martin Drive, Baltimore, MD
21218}
\affil{Electronic mail: koratkar@stsci.edu}

% The abstract environment prints out the receipt and acceptance dates
% if they are relevant for the journal style.  For the aasms style, they
% will print out as horizontal rules for the editorial staff to type
% on, so long as the author does not include \received and \accepted
% commands.  This should not be done, since \received and \accepted dates
% are not known to the author.

$^1$ Based on observations with the NASA/ESA Hubble Space Telescope obtained 
at the Space Telescope Science Institute, which is operated by AURA, Inc., 
under NASA contract NAS5-26555.
\newpage

\begin{abstract}

We present and discuss HST (WFPC2 and FOC) images and ultraviolet (GHRS) spectra
plus ground-based optical spectra of three Seyfert 2
nuclei (NGC 7130, NGC 5135 and IC 3639). These galaxies together with Mrk
477 (Heckman et al 1997) were selected on the basis of ultraviolet-brightness
from a bigger sample that comprises
the 20 brightest Seyfert 2 nuclei, with the goal of studying the
Starburst-AGN connection and the origin of the so-called
`featureless continuum' in Seyfert 2 nuclei.

The data provide direct evidence of the existence of nuclear
starbursts that dominate the ultraviolet light, and that are responsible 
for the featureless continuum in these type 2 Seyfert nuclei.
The GHRS
spectra show absorption features formed in the photospheres (SV
$\lambda$1501, CIII $\lambda$1426,1428, SiIII $\lambda$1417, and SiIII+PIII
$\lambda$1341-1344) and in the stellar winds (CIV $\lambda$1550, SiIV
$\lambda$1400, and NV $\lambda$1240) of massive stars. 
Signatures of massive stars are also clearly detected in their
optical and near-UV spectra where the high order Balmer series and HeI lines
are observed in absorption. These lines are formed in the photospheres of
O and B stars, and thus they also provide strong independent
evidence of the presence
of massive stars in the nuclei of these Seyfert 2 nuclei.
Interstellar absorption lines similar to those formed in the interstellar
medium of starbursts are also observed. They are blueshifted by a few
hundred km s$^{-1}$ with respect to the systemic velocity, indicating that
the interstellar gas is outflowing. These outflows are most likely
driven by the nuclear starburst.

These starbursts are dusty, compact, and powerful. They have sizes 
ranging from less than 100
pc to a few hundred pc (much smaller than
that seen in the prototype Seyfert 2 galaxy NGC 1068). Their UV colors
imply that they are heavily reddened (by 2 to 3 magnitudes in the UV),
and the implied bolometric luminosities are of-order 10$^{10}$ L$_{\odot}$.
The bolometric
luminosities of these starbursts are similar to the estimated
bolometric luminosities of their obscured Seyfert 1 nuclei. The data on this small 
sample suggest that more powerful AGN may be related to more powerful central starbursts.
 Comparing the HST spectra to IUE spectra obtained through
apertures with projected sizes of 3 to 11 kpc (and to IRAS far-IR data) we
estimate the nuclear starbursts account for 6 to 25\%  of the total intrinsic UV luminosity 
of the entire galaxy. 
 
\end{abstract}

% The different journals have different requirements for keywords.  The
% keywords.apj file, found on aas.org in the pubs/aastex-misc directory,
% contains a list of keywords used with the ApJ and Letters.  These are
% usually assigned by the editor, but authors may include them in their
% manuscripts if they wish.


\newpage
\section{Introduction}

According to the unified scheme of AGN, the two types of Seyfert galaxies
are the same phenomenon, but they appear as Seyfert 1 or Seyfert 2 
depending on their orientation relative to the line of sight (e.g.,
Lawrence 1991). During the last decade, sufficiently convincing
observational support  has been obtained for this scenario. One of the
strongest pieces of observational evidence for this model was the discovery of
broad recombination and FeII lines in the optical spectra in polarized
light of some Seyfert 2 nuclei  (Antonucci \& Miller 1985; Miller \&
Goodrich 1990; Miller, Goodrich, \& Mathews 1991; Tran 1995).
Furthermore, the high
excitation gas observed extending out from the nucleus with conical
or biconical morphology (Wilson, Ward, \& Haniff 1988; Tadhunter \&
Tsvetanov 1989; Pogge 1989; P\'erez et al 1989; Storchi-Bergmann,
Wilson, \& Baldwin 1992; Cappetti, Axon, \& Machetto 1997) is taken as
additional observational evidence in favor of the unified scheme. However,
some problems are still left unresolved by the unified scheme. 

One of the most outstanding of such issues is the nature of the 
optical and UV continuum in type 2 Seyfert nuclei.
In most such
nuclei, the primary contributor to the optical continuum is the light 
from an old population of stars. This light is however diluted by
an additional component in which the normal spectral features of old
stars is weak or absent. This component has traditionally been called
the `featureless continuum', on the assumption that is produced by the
AGN in some way. As we will show in this paper, this continuum is not
featureless at all!
It typically contributes 10 to 30\% of the
nuclear optical continuum in type 2 Seyfert nuclei, but is much more
conspicuous in the UV.

What is this light?
After the contribution of the old stars is removed, the
remaining optical continuum has a significantly lower fractional polarization
than the broad emission-lines, implying that the majority
can not be due to scattered light from a hidden
type 1 Seyfert nucleus (
Miller \& Goodrich 1990; Kay 1994; Tran 1995). 
In a related vein,
Terlevich, D\'\i az, \& Terlevich (1990) proposed a stellar origin for
`featureless continuum' in Seyfert 2 galaxies based in the strength of
the stellar features observed in the near-infrared. If the `featureless
continuum'
is radiation scattered into our line of sight from the obscured Seyfert 1
nucleus then it is difficult to self-consistently explain why the stellar CaII
triplet feature ($\lambda$8498, 9542, 8662 \AA) has an
equivalent width similar to normal galaxies. Cid Fernandes \& Terlevich
(1995) have also proposed that a population of young stars in the vicinity of
the nucleus is the source of the featureless continuum. Heckman et
al. (1995) have constructed an ultraviolet spectral template using IUE
spectra of 20 of the brightest Seyfert 2 nuclei. They find that only
20\% of the UV continuum emission can be attributed to a hidden Seyfert 1
nucleus. They propose that most of the ultraviolet continuum in Seyferts 2
may be produced by a reddened circumnuclear starburst, which is producing
most of the far-infrared continuum detected in Seyfert 2 galaxies. Direct
observational evidence that only a small fraction of the total UV light
detected in Seyfert 2 galaxies is emitted by a hidden nucleus comes from UV
HST images of Seyfert 2 galaxies (Heckman et al 1997; Colina et al 1997a).

The existence of a population of young stars very close to the active
nucleus has been required by several AGN models. In the `Starburst Model'
for AGN proposed by Terlevich and collaborators, the activity is produced
by a centrally concentrated burst of star formation (Terlevich \&
Melnick 1985; Terlevich et al. 1992). In the so called hybrid models,
a very compact ($r \sim 10$~pc or even smaller)
circumnuclear starburst coexists with a massive central black hole,
which is responsible for generating the ionizing continuum (Norman \&
Scoville 1988; Scoville \& Norman 1988; Perry \& Dyson 1985; Perry \&
Williams 1994).

Observational evidence for a possible connection between nuclear activity
and circumnuclear starbursts in Seyfert 2 galaxies comes from the strength of the
far-infrared continuum (Rodr\'\i guez-Espinosa, Rudy, \& Jones 1986;
Dultzin-Hacyan, Moles, \& Masegosa 1990; Pier \& Krolik 1993),
the CO 115 GHz emission (Heckman
et al 1989), the strength of the CaII triplet in absorption (Terlevich et al, 1990), the departure of the galaxy or bulge blue luminosity from
the Tully-Fisher and Faber-Jackson relationship (Whittle 1992a,b; Nelson \&
Whittle 1995), the mid-infrared 10 $\mu$m emission (Maiolino et al 1995), the ratio of the luminosity in the H band and mass of the circumnuclear region in Seyfert 2 (Origlia, Moorwood, \& Oliva, 1997), 
the diffuse radio emission around some Seyfert nuclei (Wilson 1988)
and the fraction of circumnuclear starbursts in Seyfert 2 galaxies
(Gonz\'alez Delgado et al 1997).

To understand the origin of the featureless continuum in Seyfert 2
nuclei and its relationship with circumnuclear starbursts, we have
undertaken a program that comprises high resolution UV images and UV
spectra with the HST, and ground based spectra from near UV to the near
infrared. The first results, for Mrk 477, have been presented by Heckman et
al (1997). The data provide direct evidence that the continuum (from UV
to near infrared) is dominated by a circumnuclear dusty starburst in Mrk
477. Here, we present the results for three other Seyfert 2 galaxies
(NGC 7130, NGC 5135 and IC 3639). Together with Mrk 477, these nuclei 
are among the 20 brightest Seyfert 2 nuclei. Our goal is to present
general conclusions about the AGN-Starburst connection and the origin of
the featureless continuum in Seyfert 2 nuclei. The paper is organized as
follows: In Section 2 we present the criteria used to select the
objects. Section 3 presents the observations and data reduction.
Sections 4--7 deal with the results and interpretation of the HST
images, GHRS spectra, and ground-based optical spectra. We discuss our
results and their implications in Section 8, and summarize our
conclusions in Section 9.

\section{Sample selection}

The sample of Seyfert 2 galaxies was defined by Heckman et al (1995)
based on the brightness of the Seyfert nucleus. Using the compilation of
Whittle (1992), the 30 brightest Seyfert 2 were selected using the [OIII] $\lambda$5007+4959 emission line flux
and the monochromatic flux ($\nu \times F_{\nu}$)
of the nuclear non-thermal radio source at 1.4 GHz.
All of them satisfy at least one of the two following criteria: log
F$\rm_{[OIII]}\geq$ -12.0 (erg cm$^{-2}$ s$^{-1}$) and log F$_{1.4}\geq$
-15.0 (erg cm$^{-2}$ s$^{-1}$). The sample was restricted to the 20 cases with existing IUE data. The criteria used guarantees that the sample is unbiased with
respect to the presence or absence of a nuclear starburst. The targets 
and properties are summarized in Table 1 of Heckman et al (1995). NGC 1068
was omitted from the sample used by Heckman et al to form the IUE 
spectral template, since it would otherwise have dominated the template
(which weighted spectra based on the square of their signal-to-noise
ratio).

We have taken ground-based optical  spectroscopy of all
the targets accessible with the 4m telescope at Kitt Peak National Observatory.
 HST UV images have been obtained for 12 of the 20 Seyfert 2
nuclei using the Faint Object Camera (FOC). We have also obtained
UV spectra of 4 of these 12 Seyfert 2 nuclei with the HST Goddard
High Resolution Spectrometer (GHRS). These are: Mrk 477, NGC 7130 (IC 5135), NGC
5135 and IC 3639 (Tol1238-364). They were chosen from the subsample of 12 Seyfert 2
nuclei for having the highest UV flux on arcsec scales. The results for
Mrk 477 have already been presented by Heckman et al (1997) and a summary
of the results is presented in this section. Here we present the UV
images, UV spectroscopy, and optical spectroscopy of
the other three galaxies. HST WFPC2 images, retrieved from the HST
archive, are also included. In the following subsections we present a
summary of the characteristics of these galaxies. Table 1 lists some
properties of these galaxies.

\subsection{NGC 7130}

The  NGC 7130 K band shows an inner bar
oriented
at P.A.=0$^{\circ}$ not visible at optical wavelengths  (Mulchaey et al
1997). In the digital sky survey it
can be seen that NGC 7130 has two dwarf companion galaxies located to the
North-West at 50 arcsec (15.5 kpc) and to the South-West at 30 arcsec
(9 kpc). An H$\alpha$ image shows two bright arms and circumnuclear
extended emission, but the [OIII] image only shows emission concentrated
in the nucleus (Shields \& Filippenko 1990). There is an extremely
luminous infrared source, with a very compact radio source (Norris et al
1990). It was classified as a Seyfert 
nucleus by Phillips, Charles, \& Baldwin (1983). It shows high excitation lines,
such  as [NeV] $\lambda$3426 and HeII $\lambda$4686, with line ratios
typical of Seyfert 2 nuclei ([NeV]/H$\beta$=0.78,
HeII $\lambda$4686/H$\beta$=0.16, [OIII]/H$\beta$=6.0, [NII] 
$\lambda$6584/H$\alpha$=0.96 taken from Shields \& Filippenko (1990)). The IUE
spectrum (through a large 10$\times$20 arcsec aperture, equivalent to
3.1$\times$6.2 kpc) shows 
starburst characteristics mixed with emission lines (Thuan 1984; Kinney 
et al 1991). This dual nature has also been shown at optical (Shields \&
Filippenko 1990) and NIR wavelengths (Goldader et al 1997). At optical
wavelengths, stellar absorption underlies the Balmer emission lines,
and at 2 $\mu$m, the CO stellar absorption band is clearly detected.

\subsection{NGC 5135}

NGC 5135 belongs to a group of seven galaxies
(Kollatschny \& Fricke 1989). It has a NIR bar aligned at P.A.$=123^{\circ}$
(Mulchaey et al 1997). The nucleus has also been classified as a
Seyfert (Phillips et al 1983). The nucleus shows high excitation lines with
line ratios typical of a Seyfert 2 ([NeV] is present, HeII/H$\beta$=0.18,
 [OIII]/H$\beta$=4.8, [NII]/H$\beta$=5.5 taken from Phillips et al 1983).
The spectrum 
through the IUE aperture (projected size 2.7$\times$5.4 kpc) shows mixed Seyfert
and Starburst characteristics (Thuan 1984; Kinney et al 1993). The radio
continuum map at 6 and 20 cm shows an asymmetric structure with faint
emission extended to the NE (P.A.$=30^{\circ}$) of the bright core
source. The overall linear extent is about 9 arcsec. The H$\alpha$ narrow
band image shows extended emission aligned with the radio emission
(Haniff, Wilson, \& Ward 1988;  Garc\'\i a Barreto 1996). The high
excitation gas, mapped through the [OIII] $\lambda$5007 emission line is
aligned North-South on a 2 arcsec scale. 

\subsection{IC 3639}

IC 3639 has a NIR bar aligned at P.A.=150$^{\circ}$ 
(Mulchaey et al 1997). In the digital sky survey IC 3639 has two companions: a ring galaxy
1.8 arcmin (22 kpc) to the North-East, and an edge-on galaxy 2.6 arcmin 
(33 kpc) to the North-West. The radio map at 20 cm shows a strong core
embedded in
diffuse emission, but at 6 cm only the core is detected (Ulvestad \&
Wilson 1989). The excitation ratio [OIII]/H$\beta$ is equal to 7.8
(Whittle 1992b), and conspicuous [NeV] and HeII is also present. 

\subsection{Mrk477}

Mrk 477 is in interaction with a companion located  50 arcsec 
(37 kpc) to the North-East (De Robertis 1987). 
It is the most luminous Seyfert nucleus in the Whittle compilation in terms
of [OIII], and it is known from optical
spectropolarimetry that it harbors an obscured Seyfert 1 (Tran 1995).
Our results (Heckman et al 1997) from the analysis of the UV,  and optical 
wavelength region, indicates that a compact (the effective
radius is 0.2 kpc) dusty starburst dominates the light from the UV to the
near infrared. Evolutionary population synthesis models applied to the
UV SiIV$\lambda$1400 absorption feature and to the spectral energy distribution indicate
that the star formation has occurred in Mrk 477 in a very short period of
time 6 Myr ago. This burst age is compatible with the possible presence of
$3\times10^4$ WR stars and the detection of a strong CaII triplet feature
produced by red supergiants. The bolometric luminosity of this nuclear starburst
is $\simeq4\times10^{10}$ L$\odot$, making a significant contribution to
the total bolometric luminosity of Mrk 477.

\section{Observations and Data Reduction}

\subsection{HST UV imaging}

UV images of these three galaxies were obtained as part of an HST Cycle
5 program (GO 5944), using the FOC with the F/96 relay through the filter
F210M ($\rm\lambda_c=2150$ \AA, $\Delta\lambda=212$ \AA). This filter was
selected because the bandpass does not include any strong emission lines
and has a minimal red leak; therefore, the emission is dominated by the
UV continuum at 2150 \AA. The FOC was configured in the zoomed
512zoom$\times$1024 pixel format, with a field of view of about 14$\times$14
arcsec. The integration time was 2000 s. A shorter exposure
time (356 s) was also taken in the unzoomed 512$\times$512
pixel configuration. The pixel size is 0.014 arcsec. 

The data were processed with the HST pipeline software that includes the
subtraction of a dark frame, correction for geometric distortion,
flat-fielding, and flux calibration. We have also corrected for the
non-linearity of the FOC using the algorithm of Baxter (1994) and a FOC
linearity parameter a=0.73 for the 512$\times$512 format and a=0.11 for
the 512zoom$\times$1024 format. The maximum correction factor applied to
the brightest pixels is 1.16, 1.07, and 1.05 for the
512zoom$\times$1024 format for NGC 7130, NGC 5135 and IC 3639, respectively. 
The corresponding factors for the 512$\times$512 format 
are 1.02, 1.01 and 1.01.

\subsection{HST UV spectroscopy}

In cycle 6,  UV spectra were obtained with the HST (program number
6539). The galaxies were observed with the GHRS and the G140L grating,
which has a nominal dispersion of 0.57 \AA/diode, using the Large Science
Aperture (LSA, 1.74$\times$1.74 arcsec). The acquisition of the targets
was done with the FOS blue side detector, then the LSA was pointed to the
peak of the UV surface brightness. Two different wavelength settings were
used for each galaxy, covering 1175-1462 \AA\ and 1314-1600 \AA\ for NGC
7130, 1162-1449 \AA\ and 1313-1600 \AA\ for NGC 5135, and 1162-1449 \AA\
and 1301-1587 \AA\ for IC 3639. These ranges provide about 148 \AA\ of
overlap between the spectra. The total integration times were 1461 s, 1448
s and 2898 s (split in two equal integrations) in the blue spectral
range for NGC 7130, NGC 5135 and IC 3639, respectively. In the red
spectral range, the total integration times (split into two integrations)
were 2628 s and 2652 s for NGC 7130 and NGC 5135, respectively. For IC
3639 it was 5204 s, split into four integrations. 

The instrumental spectral resolution for a point source observed through
the LSA is 0.8 \AA, but for an extended object it depends on the size of
the UV source. Using the Galactic CII $\lambda$1335 interstellar
absorption line we determine that the instrumental resolution is 
$<$ 2.8$\pm$0.5 \AA\ and $<$ 2.1$\pm$0.3 \AA\ for the blue and the red
spectra of NGC 7130, $<$ 2.7$\pm$0.2 \AA\ and $<$ 2.5$\pm$0.2 \AA\ for the blue
and red spectra of NGC 5135, and $<$ 1.9$\pm$0.1 \AA\ for IC3639 (equal for
the blue and red spectra). Thus, the instrumental resolution is of order of 500 km s$^{-1}$. Note that this CII line falls in the overlap
wavelength region of the blue and red spectra.

The center of the Galactic CII $\lambda$1335 line (assuming a gaussian
profile) is 1333.9$\pm$0.1 \AA\ and 1333.6$\pm$0.1 \AA\ for the blue
and the red spectra of NGC 7130, 1333.8$\pm$0.1 \AA\ and 1334.1$\pm$0.1
\AA\ for the blue and the red spectra of NGC 5135, and 1334.70$\pm$0.05
\AA\ for IC 3639. These give a correction to the zero point wavelength
calibration of -0.6 \AA\ and -0.9 \AA\ for NGC 7130, -0.7 \AA\ and -0.4
\AA\ for NGC 5135, and 0.2 \AA\ for IC 3639. We have corrected the
nominal wavelength scale, assuming that this apparent shift (of order of 100 km s$^{-1}$) is due to
errors in centering the targets within the LSA aperture. 

After standard pipeline processing, the spectra were combined into a
single spectrum for each galaxy covering 1175-1600 \AA, 1162-1600 \AA\
and 1162-1587 \AA\ for NGC 7130, NGC 5135 and IC 3639, respectively.

Finally, the spectra were corrected for redshift with the values given in
Table 1.

\subsection{Ground-based spectroscopy}

Longslit spectra were obtained with the 4 m Mayall telescope at Kitt Peak
National Observatory in two runs in 1996 February and October. We used the
R.C. spectrograph and the T2KB CCD chip with a spatial sampling of 0.7
arcsec/pixel. The slit width was set to 1.5 arcsec for all the
observations. NGC 5135 and IC 3639 were observed in February 16 at
paralactic angle (P.A=173$^{\circ}$, and 170$^{\circ}$ respectively)
with the grating KPC-007, covering the spectral range from 3400 \AA\ to
5800 \AA\ with a dispersion of 1.39 \AA/pixel. The integration time was
2400 s and 1800 s respectively, split in two exposures for each galaxy. 
The air mass was 2.8 and 2.1 for NGC 5135 and IC 3639 respectively.

NGC 7130 was observed in October 10 and 11 with the slit at paralactic angle using two
different gratings. KPC-007 covering the spectral range from 3400 \AA\ to
5800 \AA\ at P.A=177 $^{\circ}$. The dispersion was 1.39 \AA/pixel and the
integration time, split into two exposures, was 3000 s. The grating BL420
was used covering the spectral range from 7060 \AA\ to 10100 \AA\ with a
dispersion of 1.52 \AA/pixel. The integration time was 1800 s. The color
filter GG495 was used to block the blue light. The air mass was 2.5.

The data were reduced using the FIGARO data processing software. Bias
subtraction and flatfielding were performed in the usual way; a two
dimensional wavelength calibration was done using ARC2D (Wilkins \& Axon
1991). The spectra were corrected for atmospheric extinction using a
curve appropriate for the observatory. The frames were flux calibrated
using the standards HD 84937, BD 174708 and HD 19445 observed
during the same nights with the same set up but with a 10 arcsec slit
width. The frames were corrected for distortion along the spatial
direction by using a second order polynomial fit to the spectral
distribution of the peak continuum emission of the standard stars, and
then applied to the galaxy frames. Sky subtraction was performed in
every data frame using the the outermost spatial increments where no
obvious emission from the galaxy was present. Finally, frames of the same
wavelength range were co-added.

For subsequent analysis, we have binned the central-most 5 pixels
along the slit. Thus, the spectra we discuss in this paper pertain
to a region 1.5 by 3.5 arcsec in extent.

\section{HST optical imaging results}

Optical images of the three galaxies were retrieved from the HST archive.
The images were obtained with the HST WFPC2 through the filter F606W
($\rm\lambda_c$=6010.6 \AA, $\Delta\lambda$=1497 \AA). The  bandpass of
this filter includes strong emission lines as [OIII] and H$\alpha$. The integration
times were 500 s for each image. The pixel size of the PC, where the 
central region of these galaxies falls, is 0.046 arcsec/pixel. Figure 1
shows a global view of each galaxy. 

NGC 7130 looks very asymmetric, with a spiral arm to the North-West  being more
distorted and less organized than the arm to the South-East. This is
probably related to the presence of the dwarf companion galaxy located
to the North-West of NGC 7130. The bar seen in the NIR at
P.A=0$^{\circ}$ (Mulchaey et al 1997) is not seen in the WFPC2 image;
however, we detect two spiral arm segments located in the outer part of
two dust lanes that run in the North-South direction toward the center
(Figure 1a). These lanes could be tracing the leading edge of the NIR bar.
The inner structure is boxy with the major axis aligned at 
P.A.=90$^{\circ}$, perpendicular to the NIR bar.

The WFPC2 image of NGC 5135 shows the bar and the two spiral arms emerging from
it. Two inner
spiral arm segments located at the outer part of two dust lines are
clearly detected in the WFPC2 image (Figure 1b). These segments seem to be
spiraling to the center, and they could be located along the leading edge
of the bar.

IC 3639 clearly shows its bar at P.A.=150$^{\circ}$ and two spiral arms
emerging from the edges of the bar. This galaxy does not show obvious
signs of interaction.

In the three galaxies, the central 1 arcsec is resolved into several
structures that will be described in the next section when we compare them with the UV images.

\section{HST-UV imaging: results and interpretation}

Below we provide a detailed description of the UV morphology of the galaxies.
Some quantitative photometric properties extracted from the images are listed
in Table 2.  These include the effective (or half-light) radius $r_e$ of the
central structure, the corresponding effective surface brightness $\mu_e$,
the total flux within the $512z\times 1024$ FOC image after sky subtraction
and the total F210M flux measured from the IUE spectra as measured with the
IRAF/STSDAS SYNPHOT package.  Note that the IUE flux is typically higher than
the FOC flux often by a considerable factor (e.g.\ IC3639). This is because
the total fluxes in our images are very sensitive to the sky level which is
difficult to accurately determine.  Typically the 1$\sigma$ uncertainty in
the sky measurement integrated over the FOC area amounts to about half the
missing flux.  Sky is measured on the images far from the obvious sources of
UV emission, but low surface brightness extended emission may have been
subtracted away with the sky.  In addition since the dimensions of the IUE
are comparable or larger than the FOC image, some of the missing flux may be
in sources outside of the field of view of the images.  This uncertainty in
the total flux gives rise to a corresponding uncertainty in the ``true''
$r_e$ and $\mu_e$ of the total UV light distribution.  Instead we measure the
$r_e$ and $\mu_e$ of an inner high surface brightness component of the images
which we associate with the starburst plus Seyfert 2 nucleus.  The extent of this
component is easily discernable as an inflection point in radial growth
curves.  The inflection point is the result of a change in slope of the
surface brightness profile.  Surface brightness profiles of the three
galaxies are shown in Figure 2 illustrating that the inflection point
occurs typically at $\mu_{\rm F210M} \sim 18.5\, {\rm STmag\, arcsec^{-2}}$.
We take $r_e$ to be the radius enclosing half of the flux within this
inflection point, and $\mu_e$ to be the mean surface brightness within $r_e$.

\subsection{NGC 7130}

The highest surface brightness of the UV continuum emission is located in
the inner 1 arcsec. However, several knots are also detected along the
inner spiral arm segments that we associated with the leading edge of the
bar (Figure 3). The 1 arcsec structure at optical wavelengths (Figure 4a) is very similar
to the UV structure  (Figure 4b). The central UV continuum consists of several knots
distributed in an asymmetric ring with a total span of 1 arcsec (310 pc) in
the North-South direction, and 0.7 arcsec (220 pc) in the East-West
direction. We have taken as the origin for the coordinate system of the UV image
the point of maximum emission which is one of the knots in the ring
(Figure 4b). We have used the similarity between
them to register the images. However, the maximum in the optical image 
coincides with a weak UV knot located in the inner part of
the ring at 0.2 arcsec West and 0.1 arcsec South of the zero point.
This knot is presumably the nucleus of the galaxy. As in Mrk 477 (Heckman et al
1997) and other Seyfert 2 galaxies (Colina et al 1997), the nucleus appears to be heavily 
obscured at UV wavelengths, emitting only a small fraction of the
total light at 2150 \AA.

The total UV emission detected in our FOC frame is about 57\%\ of that
measured in the IUE spectrum. The emission from the ring is
$2.6\times10^{-15}$ erg s$^{-1}$ cm$^{-2}$ \AA$^{-1}$, or about 67\% of the
total FOC emission. The flux of the brightest knot is $7.1\times10^{-16}$ erg
s$^{-1}$ cm$^{-2}$ \AA$^{-1}$ and of the presumed nucleus is
$2.5\times10^{-16}$ erg s$^{-1}$ cm$^{-2}$ \AA$^{-1}$. The emission of this
nuclear knot thus represents less than 6\% of the total UV emission detected,
and about 10\% of the emission in the inner 1 arcsec.

The outer radius of the starburst (the inflection point in the curve of
growth) is 0.75$''$, yielding $r_e = 0.26$ arcsec (80 pc), and $\mu_e = 14.4 {\rm STmag\,
arcsec^{-2}}$.  The Galactic extinction towards NGC 7130 is E(B-V)=0.046,
derived assuming ${\rm N_{HI}/E(B-V)= 4.93 \times 10^{21}~cm^{-2}}$ (Bohlin
1975) and an HI column density of $2.27\times10^{20}$ cm$^{-2}$ (Start et al
1992). This implies an extinction of 0.45 mag at 2150 \AA. We defer our
discussion of the intrinsic extinction associated with the nucleus to section
6 below.

\subsection{NGC 5135}

As in NGC 7130, the highest surface brightness of the UV continuum emission
is in the inner 2 arcsec. This inner structure is sprinkled with about a
dozen prominent knots, much like those seen in ``normal'' starburst galaxies
(Meurer et al.\ 1995). A few fainter knots are also detected in the
spiral-arm-like segments near the center in the South-North direction and two
other knots in the North-South direction (Figure 5). The adopted origin of
our coordinate system is the knot with the hightest surface brightness
emission. However, the maximum surface brightness in the optical light 
is in the knot located at 1.05 arcsec North and 0.4 arcsec East of the zero
point (Figure 6a). This knot at the UV light seems to be surrounded by diffuse 
extended emission in the
South-West direction that comprises two other knots (Figure 6b). The symmetry of the two
inner spiral segments suggests that this knot is the center and nucleus of
the galaxy (see Figure 1b). The morphology of the inner 2 arcsec in the UV
and WFPC images is very similar. The brightest knots in the UV
seem to be formed at the end of the spiral arm segment that runs from North
to South to North.

The total UV emission detected in the $512z \times 1024$ FOC image is 70\% as
large as the flux detected by IUE at 2150 \AA. The flux detected in the inner
$2\times2$ arcsec ($540\times540$ pc) is $4.3\times10^{-15}$ erg s$^{-1}$
cm$^{-2}$ \AA$^{-1}$, which represents about 70\% of the total FOC flux. The
flux of the brightest knot is $4.6\times10^{-16}$ erg s$^{-1}$ cm$^{-2}$
\AA$^{-1}$, 7\% of the total emission. The flux of the presumed nucleus is
difficult to measure accurately due to the extended emission around it, but
it is about $2\times10^{-16}$ erg s$^{-1}$ cm$^{-2}$ \AA$^{-1}$, and amounts
to only 3\% of the total emission.

The starburst has an outer radius of 1.67$''$ yielding $r_{e} = 0.75$ arcsec
(200 pc) and $\mu_{e} = 16.1$ STmag arcsec$^{-2}$ at 2150 \AA.  The Galactic
extinction derived from an HI column density of $4.6\times10^{20}$ cm$^{-2}$
is E(B-V)=0.094. This implies an extinction of 0.92 mag at 2150 \AA\ due to
our galaxy alone.

\subsection{IC 3639}

The brightest UV emission is concentrated in the inner $0.8\times0.4$ arcsec
($170\times84$ pc). Outside of this central structure we don't find any other
knot of continuum emission in the FOC images. This central structure is
resolved into several substructures, with a morphology very
similar in the optical light (Figure 7a) and UV light (Figure 7b). Its overall $\bf{S}$ shape is aligned
perpendicular to the bar. This subarcsec structure could be
related to the inner Inner Lindblad Resonance. Here, we have taken as origin
of the plot the point with highest surface brightness in the WFPC2
image. This point is located in middle of the $\bf{S}$ structure. We suggest
that the nucleus of the galaxy is in this central structure seen in the UV
image. However, the knot of highest UV surface brightness is at 0.2 arcsec
East and 0.05 arcsec North of the nucleus.

The total FOC UV emission is only about a quarter of the IUE flux. If our sky
estimate is decreased by its $1\sigma$ uncertainty, then the total flux more
than doubles becoming $2.9\times 10^{-15}\, {\rm erg\, cm^{-2}\, s^{-1}\,
\AA^{-1}}$, or about 56\%\ of the IUE flux.  This suggests that much of the
IUE flux may result from fairly low surface brightness emission.  The central
structure emits $8.8\times10^{-16}$ erg s$^{-1}$ cm$^{-2}$ \AA$^{-1}$, this
represents 66\% of the FOC total detected emission.  This flux is emitted
mainly by the presumed nucleus ($1.3\times10^{-16}$ erg s$^{-1}$ cm$^{-2}$
\AA$^{-1}$, 10\% of the total emission) and the two extended knots of the
$\bf{S}$ ($3.3\times10^{-16}$ erg s$^{-1}$ cm$^{-2}$ \AA$^{-1}$ per knot).
The outer radius of the combined central structure is 0.51$''$ and it has
$r_{e}$ = 0.26 arcsec  (55 pc) and $\mu_{e} = 15.5$ STmag arcsec $^{-2}$. The Galactic
extinction derived from an HI column density of $6.8\times10^{20}$ cm$^{-2}$
is E(B-V)=0.14. This implies a correction of 1.36 mag at 2150 \AA\ due to our
Galaxy alone.

\section{HST-UV spectroscopy: results and interpretation}

Figure 8 shows the GHRS spectra of NGC 7130 (a), NGC 5135 (b), and IC
3639 (c). The three spectra are very similar, showing the typical
absorption features of starburst galaxies, indicating that the UV light
is dominated by the starburst component. However, they also show narrow
emission lines typical of Seyfert 2 galaxies. The contribution
of the emission lines in the spectrum of IC 3639 is stronger than in the
other two galaxies. 

\subsection{Emission line spectrum}

The most prominent emission lines are L$\alpha$ and CIV $\lambda$1549.
Other high-ionization lines such as NIV] $\lambda$1486, SiIV+OIV]
$\lambda$1397+1403 and NV $\lambda$1240, and low-ionization lines, such as CII
$\lambda$1335 are also present at least in some of the nuclei. These lines are prominent in IC 3639 and
almost absent in NGC 5135. This makes the spectrum of IC 3639 similar to
that of Mrk 477 (Heckman et al 1997). However, the absorption features are
stronger in IC 3639 than in Mrk 477. The emission lines are mainly
produced in the Narrow Line Region (NLR) of the AGN, but some
contribution to the CIV, SiIV and NV comes from the starburst component (stellar wind),
and this is probably the case in IC 3639. However, the starburst is the
dominant component in NGC 5135, where CIV and SiIV are weak and NIV] is
absent. NGC 7130 represents an intermediate case, since NIV] is absent
but CIV and SiIV are stronger than the contribution expected from the
starburst (see next section). L$\alpha$ is strong in the three galaxies.
This line can be produced by the AGN or by the starburst component. 

In IC 3639, all the emission lines CIV, NIV], NV, and L$\alpha$ show a double peak. The
observed separation
between the two peaks of CIV and NV lines are equal to the expected separation
between the two components of CIV and NV doublets, and the peaks are at
the rest wavelength of these lines. On the other hand, the two peaks in L$\alpha$ fall
170 km s$^{-1}$ redshifted and 620 km s$^{-1}$ blueshifted with respect to the systemic
velocity. The fluxes are $1.16\times10^{-13}$ and $1.4\times10^{-14}$
erg s$^{-1}$ cm$^{-2}$, and the FWHM's (not corrected for instrumental
resolution)
are 2.05 \AA\ and 1.7 \AA\, for the
red and blue components, respectively. The profiles look
asymmetric, with the blue wing of the redshifted component and the red
wing of the blueshifted component absorbed. This double peak could be
produced by absorption in outflowing gas driven by the AGN or by the starburst. 

In NGC 7130, L$\alpha$ appears asymmetric. The blue side of the profile
drops rapidly, and the peak of the line is redshifted 0.5 \AA\ (120 km s$^{-1}$) with
respect to the systemic velocity. Even though the shift is lower than the
shift observed in some starburst galaxies (Gonz\'alez Delgado et al 1998;
Kunth et al, 1997), the shape of the profile suggests that we are seeing
here the same phenomenon. The explanation is that the neutral gas
responsible for the absorption is blueshifted (outflowing)
with respect to the systemic
velocity. The flux of the L$\alpha$ emission line is $9.3\times10^{-14}$ erg
s$^{-1}$ cm$^{-2}$ and the FWHM is $3.0\pm0.1$ \AA\ (not
corrected for instrumental resolution).

In NGC 5135, L$\alpha$ is broader than in the other two galaxies, with
a $\rm FWHM=4.1\pm0.1$ \AA\ and a flux of $1.2\times10^{-13}$ erg
s$^{-1}$ cm${-2}$. The center of the line is absorbed, leaving double emission
peaks at -410 km s$^{-1}$ and 180 km s$^{-1}$ relative to the systemic velocity. Again, this is suggestive of absorption by outflowing gas.  

No broad component (width of several thousand km/s) is detected in L$\alpha$ or
CIV, indicating that scattered light from a hidden Seyfert 1 nucleus
does not contribute significantly to the UV spectrum of these Seyfert 2 nuclei.
To illustrate this we have compared the spectrum of IC 3639 (which is the
galaxy with the largest contribution from the Seyfert component, as
shown by its emission lines) with the spectrum of the average Seyfert 1 
and low redshift QSO observed by IUE. Following Heckman et al
(1995), we iteratively decrease the fractional contribution of the
Seyfert 1 spectral component to the UV continuum until the result is
consistent with the lack of broad emission in the L$\alpha$ and CIV
lines. Figure 9 shows that less than 5\% of the UV continuum can
be due to scattered light from a hidden Seyfert 1 nucleus. Because CIV in
emission is weaker in the other two Seyferts than in IC 3639, we expect
that the contribution of scattered light from a hidden Seyfert 1 nucleus
in NGC 5135 and NGC 7130 is completely negligible in the GHRS aperture. 

\subsection{Interstellar absorption lines}

The spectra of these three Seyfert 2 nuclei are very rich in absorption
features formed in the interstellar medium. The most conspicuous lines are
CIV $\lambda$1550, SiII $\lambda$1526, SiIV $\lambda$1400, CII
$\lambda$1335, OI+SiII $\lambda$1303 and SiII $\lambda$1260. Some of them
are detected at zero redshift (in the observed spectrum), and are thus 
associated with gas in the Milky Way. However, we also detect
absorption features formed in the interstellar medium of the Seyfert
galaxies. These are very similar to those observed in starburst galaxies
(Gonz\'alez Delgado et al 1998), and unlike Mrk 477,
where no interstellar absorption line associated with the Seyfert galaxy itself were detected.
Instead, in Mrk 477 the lines are observed in emission (Heckman et al 1997). The
reason is that the gas clouds in Mrk 477 seem to be illuminated by a
powerful anisotropic UV source that has the net effect of scattering
photons into our line-of-sight, leading to emission. Some of the absorption lines that we observe 
in NGC 7130, NGC 5135 and IC 3639 are superposed on, or blended with emission lines. For example, absorption in CII
$\lambda$1335 (1334.5 \AA, Morton 1991) is partially superposed on a
blend of emission multiplets that include the excited fine-structure transitions
of CII $\lambda$1335.7. 
Some of the interstellar absorption lines are also blended with wind resonance or photospheric
absorption stellar lines, for example SiIV $\lambda$1400 and CIV
$\lambda$1550.

Where are the interstellar lines formed? To address this question we need
to consider only pure interstellar lines and exclude lines such as CIV and
SiIV which are strongly contaminated by the blueshifted stellar wind features, and CII $\lambda$1335 because its radial velocity reflects
the effect of blending with the emission line CII $\lambda$1335.7. Thus, we only consider
SiII $\lambda$1260 and SiII $\lambda$1526. We measure their radial
velocity with respect to the adopted systemic velocity for each galaxy.
If we assume that the Galactic CII $\lambda$1335 is at a heliocentric
velocity of zero, then the center of the SiII lines indicates that they are
blueshifted by 720, 240, and 290 km s$^{-1}$ with respect to the systemic velocity in NGC 7130, NGC 5135 and IC
3639, respectively. This blueshift indicates that the warm interstellar gas in these
Seyfert galaxies is outflowing (consistent with our inferences based on
the L$\alpha$ emission-line).
 Outflows of similar velocity have been
detected in starburst galaxies (Gonz\'alez Delgado et al 1998; Kunth et al 1997) and in the
post-starburst galaxy NGC 1705 (Heckman \& Leitherer 1997; Sahu \& Baldes 1997). 

In NGC 7130 and NGC 5135 the interstellar absorption lines are not
spectrally resolved: their width is similar to those of the Galactic
absorption lines, indicating that the UV source in these two Seyferts is
extended and fills much of the GHRS aperture. However, the FWHM of SiII
$\lambda$1260 in IC 3639 is 2..65$\pm$0.5 \AA, corresponding to an intrinsic
FWHM of 440$\pm$120 km s$^{-1}$. This broadening of the lines is more
evidence of the large scale motions of the interstellar gas in IC 3639.

Further evidence of the large scale motions of the interstellar gas comes
from the equivalent widths of the SiII lines, which have values between 1-2
\AA. These equivalent width are
similar to those observed in starbursts
(Leitherer et al 1996; Conti et al 1996; Gonz\'alez Delgado et al 1998;
Heckman et al 1998) and larger than the interstellar lines observed in the spectra of nearby stars.
They are optically thick, because SiII $\lambda$1260 and SiII
$\lambda$1526 have similar equivalent widths. This means that they are in
the flat part of the curve-of-growth, and their equivalent width is
mainly related to the velocity dispersion of the gas instead of the
column density. Therefore, their equivalent width implies a velocity
dispersion of order of 150 km s$^{-1}$ for NGC 7130 and NGC 5135, and 120 km
s$^{-1}$ for IC 3639.

\subsection{Photospheric absorption lines}

The spectra of these Seyferts are also rich in absorption features formed
in the photospheres of massive stars. The most prominent lines are SV
$\lambda$1501, CIII $\lambda$1426,1428, SiIII $\lambda$1417 and SiIII
$\lambda$1341-1343+PIII $\lambda$1344. These lines are strong in late O and early B stars
(Walborn, Bohlin, \& Panek 1985; Walborn, Parker, \& Nichols
1995). They have been detected in the prototype starburst galaxy NGC 7714
(Gonz\'alez Delgado et al 1998, in preparation). These are not
resonances lines, and therefore they cannot be formed in the interstellar
medium. The measured equivalent widths of these lines are given in Table 3.
Their strength indicates that O and early B stars are the dominant
population in the UV spectrum of these Seyfert galaxies. 

Another possible feature that could be associated with the starburst is
the photospheric lines SiIII $\lambda$1294,1296. These lines are very
strong in early B supergiants. We note in our UV spectra that the
absorption feature around 1300 \AA\ is stronger and broader than the other
interstellar absorption lines, indicating that the interstellar lines
OI+SiII $\lambda$1303 could be contaminated by SiIII. These SiIII lines
have been detected in the super star-cluster in NGC 1705 (Heckman
\& Leitherer 1997).

\subsection{Wind absorption lines}

The other absorption features stronger than the photospheric lines seen in
the spectra are the resonance lines CIV $\lambda$1550, SiIV $\lambda$1400
and NV $\lambda$ 1240. They can form in the photospheres of O and B
stars; however, they develop a P Cygni profile if they form in the
stellar wind of O stars. In that case, the absorption feature shows a
velocity blueshift of the order of a thousand km s$^{-1}$. These lines are
very prominent if the spectrum is dominated by a starburst with age
between 1 to 10 Myr. The profiles of these lines have been used to derive
the stellar content and star formation history in starburst galaxies
(Robert, Leitherer \& Heckman 1993; Leitherer et al 1996; Conti et al
1996; Gonz\'alez Delgado et al 1998). The UV spectrum of these Seyfert 2 galaxies 
is very similar to the starburst galaxy NGC 1741 (Conti et al 1996). This
similarity suggests that the stellar content in these Seyfert
galaxies is similar to typical starburst galaxies.

To prove this conclusion we use the line profile synthesis technique
for the wind lines. Evolutionary synthesis models of a stellar population
for these lines has been computed by Robert et al (1993) and Leitherer et
al (1996) based on an IUE library of O and B stars. The models assume
solar metallicty, and the lines are synthesized for different star
formation histories (instantaneous and continuous star formation) and
different assumptions for the IMF.

The wind line least affected by the AGN contribution is SiIV, so we now use
only this line for the fit. However, later we will check that the best
solutions found with SiIV are compatible with the absorption part of the
wind profiles of the CIV and NV lines. Because of a possible contribution
from the Seyfert nebular emission to the SIV, to perform the fit we used
 only a wavelength
window from 1381 \AA\ to 1387 \AA\ on the blue side of the wind profile. To
constrain the solutions we have computed the $\chi^2$ parameter between
the observations and every model as a function of the age, IMF
upper mass cut-off (M$\rm_{up}$),
and IMF slope for instantaneous burst or continuous star formation (csf)
models. The results are plotted for NGC 7130 (Figure 10), NGC 5135
(Figure 11) and IC 3639 (Figure 12) as bubble diagram, where the 
size of the bubble is proportional to $\chi^2$. The smallest deviations
between observations and models are given by the smallest bubbles.

SiIV shows a strong P Cygni profile if the starburst is dominated by O
supergiants. A conspicuous profile indicates that the star formation
occurs in a short period of time, due to the fact that the relative number
of these
stars with respect to the total O stars is low because their relative life time is
short. Therefore, the SiIV line is very useful to discriminate between models of 
instantaneous bursts and continuous star formation (csf).

Figure 10 shows clearly that NGC 7130 has experienced a burst of star
formation. The best fit indicates that stars more massive than 60
M$\odot$ are present in the burst (Figure 10a), and the mass distribution
follows a Salpeter ($\alpha$=2.35) or flatter IMF (Figure 10b). The
best fits indicate an age of 3-4 Myr.

In NGC 5135, the fit favors burst models with M$_{up}\geq40$
M$\odot$ (Figure 11a), Salpeter or steeper ($\alpha$=3.0) IMF (Figure
13b), and age 3-5 Myr. However, we cannot exclude csf models
with M$_{up}\geq60$ M$\odot$ (Figure 11c), and Salpeter or flatter IMF's
(Figure 13d). 

As for NGC 5135, the SiIV line profile in IC 3639 is compatible
with both burst and csf models. Burst models with M$_{up}=60$ M$\odot$ and
age 3-4 Myr and csf models with star formation that began 4-5 Myr ago provide 
good descriptions of the data.

The results show clear evidence that massive stars are present in the
inner 1.74$\times$1.74 arcsec$^2$ (few hundred pc) of these Seyfert 2 galaxies.
The stellar
content and age of the burst in these nuclear starbursts seem to be
similar. However, we cannot exclude the possiblity that continuous star
formation is taking place in NGC 5135 and IC 3639. The UV continuum
morphology of NGC 5135 (see Figure 6b) suggests that the different knots
seen in the FOC image are super-star clusters.
If this is correct, then it favors a multiple burst scenario.

\subsection{Extinction Estimates and Intrinsic Luminosities}

We have demonstrated in the previous sections that the UV light in these
Seyfert 2 galaxies is dominated by the starburst component. Thus, the
continuum flux distribution can be used to estimate the reddening. The fluxes 
at 1500 \AA\ measured in the GHRS aperture are 3.3$\times$10$^{-15}$ 
erg s$^{-1}$ cm$^{-2}$ \AA$^{-1}$, 4.7$\times$10$^{-15}$ erg s$^{-1}$ cm$^{-2}$ \AA$^{-1}$ and 1.65$\times$10$^{-15}$ erg s$^{-1}$ cm$^{-2}$ \AA$^{-1}$ for NGC 7130, 
NGC 5135 and IC 3639, respectively. These fluxes represent the 38\%, 49\% and 
15\% of the IUE flux in the 1432-1532 \AA\ band measured by Kinney et al (1993). 
Leitherer \& Heckman (1995) have shown that the UV energy distribution
arising from a starburst has a spectral index ($\alpha$) in the range -2.6 to -2.2
(F$_{\alpha}\sim \lambda^{\alpha}$ if the starburst is younger than 10 Myr (see also Meurer et al 1995,1997)). This spectral index is
independent of the metallicity and IMF. Therefore, any deviation from
the predicted spectral index can be attributed to reddening. 

First, the spectra are corrected for Galactic extinction using the MW
extinction law and 
the Galactic  E(B-V). Then, the corrected spectra are further dereddened
to correct for the internal extinction in the starburst. This is done using
the empirical Calzetti et al (1994) 
extinction law,  such that the slope of the 
corrected spectrum (log  F$_\lambda$ vs log $\lambda$)
matches the slope 
of the synthetic spectra that fit the wind absorption lines.
Figures 14-16 show the spectra corrected by reddening and one of the best
models that fits the SiIV line. These models also fit the absorption
part of the profiles of CIV and NV well. Note also that there is good agrement
between the synthetic photospheric absorption lines (SV, CIII and SiIII) and
the observations.
 
The resulting values of the 
reddening are listed in Table 4 and the extinction-corrected UV luminosities
are given in Table 5. The typical total UV extinctions are 2 to 3
magnitudes, and the implied extinction-corrected bolometric luminosities
are of-order 10$^{10}$ L$_{\odot}$.
Note that these values pertain only to the light within the relatively small
GHRS aperture. We have followed an analogous procedure to determine the
extinction and extinction-corrected UV luminosities for the IUE spectra
(taken through an aperture about 70 times larger in projected area). These
values are also listed in Tables 4 and 5. The bolometric luminosities
implied by the extinction-corrected IUE spectra are typically a factor
of 3 to 10 larger than for the GHRS spectra, and are quite similar to
the total IR luminosities measured by IRAS (see Table 1 and 5).

In the previous section, we have seen that the line profile synthesis
technique has not provided a unique solution. Instead, we have obtained a
few models which are compatible with the profile of the wind lines. To
better constrain the solution, we can use the UV continuum luminosity at
1500 \AA. With this luminosity, we can predict the number of ionizing
photons, Q, for each model, and the mass of the burst. The predicted
ionizing photons can be compared with the values derived from the Balmer
recombination lines. Table 5 gives the
mass of the nuclear starburst and its predicted
ionizing luminosity, based on the extinction-corrected UV luminosity
(as measured through the GHRS aperture).

\section{Ground-based near-UV through near-IR spectroscopy}

\subsection{High-order Balmer series and HeI absorption-lines}

The detection of stellar features in the optical spectra of starburst
galaxies is a difficult business. The reason is that the optical spectra
of hot stars are dominated by absorption lines of H and He, with only
very weak metallic lines (Walborn \& Fitzpatrick 1990). In starbursts,
the H and He absorption features are coincident with the nebular emission
lines that mask any absorption signatures. However, the high-order
Balmer series are detected in absorption in many starburst galaxies (e.g.
NGC 7714, Gonz\'alez-Delgado et al 1995) or even in the spectra of giant
HII regions (e.g. NGC 604, Terlevich et al 1996). These features can be
seen in absorption because the strengths of the Balmer emission lines
decrease quickly with decreasing wavelength, while the stellar absorption lines have
nearly constant equivalent widths. Thus, the net effect is that
H$\alpha$, H$\beta$ and H$\gamma$ are seen in emission while the
high-order lines such as H8 to H12 can be seen in absorption.

The spectra of the three Seyferts show clearly the high order lines of
the Balmer series in absorption, H9 (3835 \AA), H10 (3798 \AA), H11 (3771
\AA), H12 (3750 \AA) (Figure 16). H$\beta$ to H$\epsilon$ also show some
absorption wings. The spectra also show HeI lines in absorption, the
stronger ones are at $\lambda$4921, $\lambda$4387, $\lambda$4143, 
$\lambda$4026 and $\lambda$3819. Note that one of the
strongest lines, HeI $\lambda$4471, is not detected in absorption. The
reason is that the corresponding nebular line is also very strong in
emission (see for example the emission line spectrum of the Orion nebula
published by Osterbrock, Tran, \& Veilleux, 1992) diluting the absorption feature.
Note also that the spectra of A stars are rich in HI lines, which are even
stronger and wider than in B stars, but they do not show any He lines in
absorption. Thus, the detection of these HeI absorption lines is strong
evidence for the presence of very hot, massive, and young (O and early B)
stars in these Seyfert 2 nuclei.

However, the spectra also show features from an old population, incluiding the G
band (4284-4318 \AA, as defined by Bica \& Alloin 1986) and the MgI+MgH
feature (5156-5196 \AA, as defined by Bica \& Alloin 1986). These are strong
in G and K stars, with equivalent widths of around 6 to 7 \AA. These features
come from the bulge component of these galaxies that contributes to the
optical light. In fact, a simple combination of a B and G star can
reproduce most of the stellar features seen in the nuclear spectra 
of these Seyferts (Figure 17). The equivalent width of MgI+MgH is 1.8 \AA,
1.9 \AA\ and 4.0 \AA, and the equivalent width of the G band is 1.9 \AA,
1.9 \AA, and 3.7 \AA, for NGC 7130, NGC 5135 and IC 3639 respectively.
These values are weaker than in K stars and also weaker than in the normal
nuclei of early-type spirals galaxies (e.g. the equivalent width of the G band is
6$\pm$1 \AA, Heckman, Balick, \& Crane 1980). From the dilution of the 
MgI+MgH feature and the G band we infer that 30\% (in NGC 7130 and NGC 5135)
and 60\% (in IC3639) of the blue light in our optical spectra comes from the old bulge stars.

As part of this program we have also observed, to be used as templates,
three galaxies that are known to be in different evolutionary
stages. They are NGC 4339, NGC 205, and NGC 1569. The spectrum of NGC 4339
shows the features typical of an old bulge stellar population (Ho,
Filippenko, \& Sargent 1995). Its spectrum is well fitted by a K0III star.
NGC 205 is a galaxy that shows the typical signatures of an intermediate
age population. Population synthesis of the optical continuum (Bica
\& Alloin1990) indicates that the optical light is dominated by a
population of stars with age 0.1-1 Gyr. The spectrum is
well described by a combination of A2I and G0V Stars. NGC 1569 is an
example of a galaxy with very young stars. Our evolutionary synthesis analysis
(Gonz\'alez Delgado et al 1997) indicates that O and B stars are the
dominant contributors to the optical light. The spectrum of a B0V star
combined with a small fraction of the light from a G0V star reproduces
well most of the stellar features in the spectrum of the cluster B in NGC
1569. 

We have compared the normalized spectra of the Seyferts with our template
galaxies. The stellar features in NGC 7130 and NGC 5135 are very well
matched by NGC 1569 (Figure 18). The stellar features of the spectrum of IC
3639 could be fitted by the spectrum of NGC 205; however, we note that
after subtracting NGC 205 from IC 3639 the resulting spectrum shows all
the Balmer lines apparently in emission with broad wings. This indicates that the Balmer
absorption lines in IC 3639 are narrower than in NGC 205, and the
population responsible for the H absorption features is probably dominated more by B stars than by A stars. We have measured the width of 
H$\delta$ in the stars from the atlas of Jacobi, Hunter, \& Christian
(1984), and the FWHM of a A2V star is 23 \AA\ and B0V star is 11 \AA.
Thus, the broad wings in the Balmer emission lines results from an
oversubtraction of the absorption features. In fact, these wings do not
show up if we subtract from IC 3639 a combination of the spectrum of NGC
1569 with NGC 4339; 60\% of the continuum light is accounted for by NGC
1569 (starburst) and 40\% by NGC 4339 (old stars), as seen in Figure 19.

Our conclusion is that massive stars (O and early-B) are the responsible for
producing the high-order Balmer series and the HeI absorption lines in the
spectra of these Seyfert 2 galaxies. However, a larger contribution is made by
the
old population in IC 3639 compared to NGC 7130 and NGC 5135. The strong nebular emission lines in IC 3639 likewise conspire against the detection of some
weaker stellar signatures of massive stars at optical wavelengths that
are more clearly detected in NGC 7130 and NGC 5135.

In view of the relative ease with which our optical spectra 
directly demonstrate the
presence
of hot stars in the nuclei of these three Seyfert galaxies, it is worthwhile
considering why the `featureless continuum' in Seyfert nuclei was ever given
such an (apparently) inaccurate name. We believe there is a reasonable
explanation for this. The three galaxies discussed in this paper
are (along with Mrk 477) the Seyfert 2 nuclei with the
highest UV fluxes (thereby enabling spectroscopy with GHRS). This means
that these galaxies also have {\it optical} `featureless continua' that are
unusually bright relative to both the old underlying stellar population
and the nebular emission (line and continuum). Thus, rather than
simply being a mysterious and rather minor contributor to the optical
light, the `featureless continuum' actually dominates the near-UV light, and can
then be seen to be anything but `featureless'. This leaves unanswered
the question as to whether the continuum  generically
has a starburst component in all type 2 Seyfert nuclei. We will return
briefly to this issue in section 8 below.

\subsection{The CaII triplet}

In the red, one of the clearest features from massive stars is the
CaII
triplet. These absorption lines are observed in most starburst
galaxies (Terlevich, D\'\i az, \& Terlevich 1990; Storchi-Bergmann, Kinney
\& Challis 1995; Gonz\'alez-Delgado et al 1995) and the spectra of young
star clusters (Bica, Alloin \& Schmidt 1990; Gonz\'alez Delgado et al
1997). These features are very strong in starbursts with ages between 10 and 
20 Myr (Mayya 1997; Garc\'\i a-Vargas, Moll\'a, \& Bressan 1997) due to
the evolution of massive stars to the red supergiant phase. This feature
is also detected in a significant fraction of Seyfert galaxies (Terlevich,
D\'\i az, \& Terlevich 1990; Nelson \& Whittle 1995; Gonz\'alez Delgado
1995), and it is as strong as that detected in the central region of
normal galaxies (where it is primarily
produced by an old population of red giants).
The non-dilution of these features  points against the
interpretation of the spectra of AGN as the sum of the spectra from an
old population plus a power law due to the AGN. Otherwise the
non-dilution of the CaII triplet can only be explained assuming a 
cut-off in power law at near-IR wavelengths (Malkan \& Filippenko 1983).
Therefore, the most natural interpretation is the presence of a starburst
population of massive stars. 

Of our targets, only NGC 7130 was observed in the red wavelength
region. The CaII triplet is clearly detected. The equivalent width
of the strongest line (8542 \AA) is 3.3$\pm$0.4 \AA\ and the weakest one
(8498 \AA) is 1.4 \AA. The line CaII $\lambda$8662 is affected by a bad
column in the CCD and its equivalent width is very uncertain. The
equivalent width of CaII $\lambda$8542 is comparable to the typical
values of this line (3 to 4 \AA) in normal early type galaxies (Terlevich,
D\'\i az, \& Terlevich 1990). 

\subsection{Optical colors}

In section 6.1 we have shown that the optical continua of our Seyfert galaxies
 are rich in H and
He I absorption lines that cannot be explained by an old stellar
population typical of the central region of normal early type galaxies.
This is a strong evidence that the optical continua of these Seyfert 2
nuclei cannot be fit by only the sum of the old bulge stellar
population plus a power law due to the AGN. Now, we check that the
continuum colors are compatible with a young stellar population if we
subtract the contribution of the old 
bulge population of stars. To subtract the old stellar component from the spectra of
these Seyferts, we use the equivalent width of
the MgI+MgH and the G band features. Comparing the equivalent width
of these signatures to the spectrum of NGC 4339, we find that the old
component contributes 30\% of the light at 4700 \AA\ in NGC 7130
and NGC 5135, and 60\% in IC 3639. We have subtracted the old component
from the spectra after correcting these with the reddening derived from
the UV continuum slope, and then we have computed the U, B and V 
magnitudes excluding the contribution from the emission lines. The colors derived are
U-B=-0.88, -0.91, and -0.85, and B-V=0.19, -0.01 and -0.14 for NGC 7130, NGC
5135 and IC 3639, respectively. These colors are compatible with an
instantaneous burst with age between 4 and 8 Myr (see figures 13 and 15 in
Leitherer \& Heckman 1995). Because the colors depend of the reddening correction aplied 
to the continuum, we have made an approximate estimation of the uncertainties associated to the colors
correcting the spectra by a color excess equal to 0.3 instead of the values listed in Table 4. It is
the largest value that we have estimated for the reddening comparing the observed UV continuum slope with the synthesized continuum. In this case, the colors derived are U-B=-0.93, -1.04, and -0.91, and B-V=0.11, -0.1, and -0.24 for NGC 7130, NGC 5135 and IC 3639, respectively. These colors are also compatible with an instantaneous burst with age between 3 and 8 Myr.  
However, the uncertainty in the determination 
of these colors is probably large because we have assumed that the old stellar 
population is affected by the same amount of extinction that the young stars.

\subsection{Hydrogen recombination lines}

In these Seyfert 2 galaxies the nebular HI recombination lines can be excited
by ionizing
radiation from hot stars associated with the starburst and also by the
active nucleus. We can estimate the contribution of the starburst to the
emission lines by comparing the number of ionizing photons predicted by the
UV continuum with that derived from these recombination lines. 

The reddening of the gas is derived adding 1-1.5 \AA\ to the observed equivalent width of the Balmer lines in emission (H$\beta$, H$\gamma$ and H$\delta$), to account for the contribution of absorption by the underlying stellar population. Then, we fit a slope to the dependence of the ratio of corrected measured flux to the theoretical value versus the reddening curve. Assuming Case
B recombination at T=10$^4$ K the E(B-V) obtained is 0.75, 0.67, and 0.52 for
NGC 7130, NGC 5135 and IC 3639, respectively. These values of the color
excess are systematically larger than those derived from the UV continuum
slope (see
Table 4). This discrepancy between the extinction derived from the two
methods has been also found in starburst galaxies (Fanelli, O'Connell, \&
Thuan 1988; Calzetti, Kinney, \& Storchi-Bergman 1994). The log of the H$\beta$
luminosity measured in the central 1.5$\times$3.5 arcsec in each spectrum
(corrected for the reddening derived from the Balmer decrement) is 41.12
(erg s$^{-1}$), 41.08 (erg s$^{-1}$), and 40.62 (erg s$^{-1}$) for NGC
7130, NGC 5135 and IC 3639, respectively. These luminosities represent a
lower limit to the total luminosity, since the ionized gas is more
extended than the spectrograph slit. The required number of ionizing photons is
$2.7\times10^{53}$, $2.5\times10^{53}$, and $0.9\times10^{53}$ ph s$^{-1}$ for
NGC 7130, NGC 5135 and IC 3639, respectively. 
It is also possible that there
is a more heavily-extincted (dustier) component of ionized gas associated
with the starburst and/or AGN. This may be more apparent at near-IR
wavelengths, so we have used the spectra of the Br$\gamma$ emission-lines
measured by Goldader et al (1997). In the cases of NGC 7130 and NGC 5135,
the implied ionizing photon luminosities are 
$3.4\times10^{53}$ and $4.5\times10^{53}$ ph s$^{-1}$ (larger than the values
based on the Balmer lines by a factor of $\sim$1.5). Comparing
the ionizing photon luminosities derived from the HI recombination lines
to those estimated for the starburst (based on UV continuum luminosity),
we find that the starburst itself can account for most of the 
Hydrogen-ionizing photons (but see Section 7.5
below). This information is summarized
in Table 5.

The equivalent width of the Balmer recombination emission lines is another
indicator of the age of the burst (Copetti, Pastoriza, \& Dottori 1986).
We have measured the equivalent width of H$\beta$ in the resulting
spectrum after subtracting the contribution of the old bulge component
and after having corrected both the line and continuum for extinction
(see Table 4). The corrected
values are 138 \AA, 110 \AA, and 131 \AA\ for NGC 7130, NGC 5135 and IC
3639, repectively. These equivalent widths are compatible with an
instantaneous burst of about  4 - 5 Myr (Leitherer \& Heckman 1995). 
However, the
values are somewhat difficult to interpret. On the one
hand, they represent a lower limit to the real
equivalent width since the ionized gas is likely to be more 
spatially-extended than the
continuum. On the other hand, they represent an upper limit, since the
hidden type 1 Seyfert nucleus will also contribute to the excitation
of these lines.

\subsection{Emission line spectrum}

Here we present only the results of the emission line spectrum that could
be related to the starburst activity in these galaxies. However, we have to
first to point out that the nuclear spectra show 
strong high-excitation emission lines
such as [NeV] $\lambda$3426 and HeII $\lambda$4686, and line ratios that are
typical of 
type 2 Seyfert nuclei. In particular, creating the NeV ion requires a copious
supply of soft X-rays (E $>$ 0.1 keV), and this rules out an ordinary
population of hot (O) stars as the only photoionization source for the
emission-line gas. After correcting by reddening, the ratio [NeV] $\lambda$3426/H$\beta$ measured in our spectra is 0.6, 0.4, and 0.7 for NGC 7130, NGC 5135 and IC 3639, respectively.

NGC 7130 and NGC 5135 show evidence of their composite spectra in the
profile of the H$\beta$ emisssion line. H$\beta$ is resolved into two
components. In NGC 7130 the narrow and the broad components have FWHM's of
4.8 \AA\ ($\simeq$300 km s$^{-1}$) and 15.8 \AA ($\simeq$960 km
s$^{-1}$) and fluxes of 1.35$\times$10$^{-14}$ erg s$^{-1}$ cm$^{-2}$ and 8.2$\times$10$^{-14}$ erg s$^{-1}$ cm$^{-2}$, respectively. The broad component is blueshifted by $\simeq$200 km s$^{-1}$
with respect to the narrow one, which is at the systemic velocity. These results are in agreement with those
reported by Shields \& Filippenko (1991). We find a similar result in NGC
5135. The FWHM of the two components of H$\beta$ is 4.3 \AA\ (260 km
s$^{-1}$) and 10.4 \AA\ (640 km s$^{-1}$), with the broad component
blueshifted by 170 km s$^{-1}$ with respect to the narrow one, which is also blueshifted by 65 km s$^{-1}$ with respect to the systemic velocity (Figure
21a). In IC 3639, H$\beta$ is well fitted with only one component with
FWHM of 5.2 \AA\ (320 km s$^{-1}$). Shields \& Filippenko (1991) suggest
that the narrow component of H$\beta$ in NGC 7130 is associated with the
starburst component and the broad one with the AGN. This latter component
would have to be associated with the so-called Narrow-Line Region of the
AGN, since
 the widths of
the broad components we observe is much lower than the width of Balmer lines
formed
in the Broad Line Region
Thus, we don't believe that it could be associated with the hidden Seyfert 1
nucleus.

[OIII] $\lambda$5007 is also resolved in two components in NGC 7130 and
NGC 5135. IC 3639 is fitted with only one component of FWHM of 7.6 \AA\
(450 km s$^{-1}$). The two components in NGC 7130 have FWHM of 19.4 \AA\
(1160 km s$^{-1}$) and 6.9 \AA\ (410  km s$^{-1}$), with the narrow 
one blueshifted by 70 km s$^{-1}$ with respect to the broad one which is at
the systemic velocity. The fluxes are 4.1$\times$10$^{-14}$ erg s$^{-1}$ cm$^{-2}$ and 9.2$\times$10$^{-14}$ erg s$^{-1}$ cm$^{-2}$ for the narrow and broad component, respectively. In NGC 5135 the FWHM of the two components is 4.4
\AA\ (260 km s$^{-1}$) and 10.3 \AA\ (610 km s$^{-1}$), with the broad
one blueshifted by 4.2 \AA\ (250 km s$^{-1}$) with respect to the
systemic velocity (Figure 21b). The fluxes are 
9.2$\times$10$^{-14}$ erg cm$^{-2}$ s$^{-1}$ and 7.7$\times$10$^{-14}$ erg
cm$^{-2}$ s$^{-1}$ for the narrow and broad components, respectively. 

HeII $\lambda$4686 in NGC 7130 and NGC 5135 is also resolved into two
components, but not in IC 3639. In NGC 7130, we can not fit the line properly
due to the absorption features close to 4686 \AA\ that make the position of
the continuum very uncertain. Consequently,
the fluxes and FWHM's can not be estimated to better than in a factor of two.
However, it is clear that HeII shows a broad component. In NGC 5135, the
broad component of FWHM 8.2 \AA\ (525 km s$^{-1}$) is blueshifted by 2.5 \AA\
(150 km s$^{-1}$) with respect
to the narrow one (FWHM $=$ 3.7 \AA). The fluxes are 2.2$\times$10$^{-15}$
erg cm$^{-2}$ s$^{-1}$
and 3.5$\times$10$^{-15}$ erg cm$^{-2}$ s$^{-1}$ for the narrow and broad
components respectively.
  
Thus, the blueshifted component in H$\beta$, [OIII] and HeII $\lambda$4686 could represent
high velocity gas in the Narrow Line Region, as seen in other Seyfert
nuclei (Wilson \& Heckman 1985 and references therein), or an outflow of
gas resulting from the star formation activity in the nuclei of these
two galaxies. Outflowing gas is also a very common phenomenon in starburst
galaxies (Heckman, Armus, \& Miley, 1990).

\section{Discussion and implications}

\subsection{Constraints on the starburst population}

The data discussed here and those for Mrk 477 (Heckman et al 1997)
clearly demonstrate that hot, massive
stars (O and B) are present in the nuclei of these Seyferts. These
starbursts are very compact. The effective (half-light) radii are
of-order 10$^{2}$ parcsecs, from about 55 pc in IC 3639 to 200 pc
in NGC 5135. They are therefore about an order-of-magnitude more compact
than the circumnuclear
starburst in NGC 1068 (Bruhweiler et al 1991) and the circumnuclear rings
reported by Colina et al (1997) in other type 2 Seyfert galaxies, which are
several kpc in diameter. 

The UV absorption features due to stellar winds
(SiIV, CIV, and NV) suggest that the duration of the starburst was short.
Is it possible that the starburst formed in an instantaneous event of star
formation? Assuming a typical velocity dispersion for these galaxies of
200 km s$^{-1}$ and a diameter of the nuclear
starburst of 330 pc, 160 pc, 405 pc and
110 pc for Mrk 477, NGC 7130, NGC 5135 and IC 3639, respectively, the
crossing times within the starburst are 1.6 Myr, 0.8 Myr, 2.0 Myr and 0.5
Myr. Thus, arguments of causality show that it is dynamically possible
for these to be effectively instantaneous bursts of star-formation
(i.e. the duration of the star-formation could be significantly
less than the evolutionary lifetime of a single generation of massive stars).
However, it is remarkable how close to the limit our derived durations
for the starburst are; the star formation that took place must have been
extremely efficient.

Our comparison with evolutionary synthesis models indicates
that we are seeing a starburst
in a particular evolutionary state. The age estimated is between 3 and 5
Myr for NGC 7130, NGC 5135 and IC 3639, and 6 Myr for Mrk 477. The equivalent
widths of the Balmer recombination lines are also compatible with a burst
of  5 Myr. This age range represents a special phase of the
star formation event, which is when the most massive stars evolve to the
Wolf Rayet (WR) phase. For solar metallicity and Salpeter IMF, it is
expected that a population of WR stars should be present in the starburst. In
Mrk 477, a possible WR signature was detected in the optical spectrum. The
feature,
emission underlying broad HeII $\lambda$4686, has a luminosity compatible with
the presence of $3 \times 10^4$
WR stars. We detect HeII in the other three Seyfert
nuclei, but the width indicates that this is likely a nebular line. Nebular
HeII has been detected in high-excitation HII galaxies (Campbell, Terlevich,
\& Melnick 1986; e.g. IZw18 Izotov et al 1997) and giant HII regions
(e.g. NGC 2363 Gonz\'alez-Delgado
et al 1994) and  appears to be excited by WR stars 
(Schaerer \& Vacca, 1997). However, the ratio HeII/H$\beta$ ($\simeq$0.02) 
in these star-forming objects is much lower than that observed in our
three Seyfert 2 nuclei (0.14, 0.16 and 0.13 for NGC 7130, NGC 5135 and IC 3639,
respectively). This suggests that the HeII nebular emission (like the [NeV]
emission) is excited by the hard radiation produced by the hidden type 1
Seyfert nucleus. We
do not detect any other WR signatures in the  optical, such as NV
$\lambda$4604,4620 or NIII $\lambda$4634,4640. Another WR star signature in
the UV region is the HeII $\lambda$1640 stellar wind line.
Unfortunately, our GHRS
spectra do not include this wavelength. However, Thuan (1984) detects
this line in the IUE spectrum of NGC 7130 and NGC 5135, and he attributes
it to WR stars.

Could the WR signatures in the optical region be diluted by a previous 
star formation event that is now in the post-starburst phase? If the
contribution of this population to the optical spectrum were larger than to the UV, the
optical signatures could be diluted significantly, but not the UV ones.
The detection of the CaII triplet could support this argument. Evolutionary
synthesis models predict that CaII is very strong in absorption if the star
formation event is in the post-starburst phase with an age between 10 and 20
Myr (Garc\'\i a-Vargas et al 1997), due to the evolution of the massive
stars to the red supergiant phase. However, for ages less than 7-8 Myr
the number of red supergiants in the starburst is too small to produce
significant CaII absorption. Thus, the strength of the CaII triplet in NGC 7130
would indicate a burst of 10-20 Myr. However, for these ages, the
stellar wind absorption lines in the UV could not be formed, because these
lines are formed only in the stellar winds of O stars. To reconcile these
two features, we require continuous star-formation,
or a significant contribution to CaII by bulge stars, or a burst with a finite time duration 
or a model of two bursts. In this last case, a young burst (3-5 Myr) to
explain the UV features and an older one (10-20 Myr) to explain the
CaII triplet is required. A similar solution was found to explain the
WR features and
the CaII triplet detected in the super stellar cluster A of NGC 1569
(Gonz\'alez Delgado et al 1997). On the other hand, the UV morphology of these starbursts (see Figure 4b, 6b and 7b) suggests that the knots detected could be super star clusters. They could be formed in  multiple bursts with different ages. Probably, the UV continuum and the stellar absorption features are dominated by the youngest cluster, and the near-infrared continuum and the CaII triplet absorption lines by the oldest one. The spread in age of the clusters is determined by the time duration of the burst. Thus, the integrated light  of the starburst could be dominated by O massive stars at the UV wavelengths and by cooler stars at the red and near-infrared.

\subsection{Starburst vs AGN Energetics}

From the UV luminosity of the GHRS spectra and the reddening derived from
the UV continuum we have estimated the mass of the burst and the
bolometric luminosity (see Table 5). The latter are typically in the range of
10$^{10}$ L${\odot}$ for the three nuclei studied here, while
the bolometric luminosity of the starburst nucleus of
Mrk 477 is a factor 3-10 larger.

How large is the starburst contribution to the total intrinsic UV luminosity
of these galaxies? Assuming that the infrared luminosity computed with the
four IRAS bands (see Table 1) is due entirely to reprocessing of the total intrinsic 
UV luminosity of the galaxy, and comparing with the bolometric luminosity of the 
central starburst as deduced from their UV luminosity via the models, 
we estimate that the central starburst
contributes about 25\% in Mrk 477, 6\% in NGC 7130, 7\% in NGC 5135,
and 6\% in IC 3639
to the total intrinsic UV galactic luminosity (see Table 5).
It is also possible (see the next paragraph) that much of the
infrared luminosity is due to reprocessing of light generated by the AGN.
In contrast, the intrinsic UV luminosities implied by the de-extincted
IUE spectra (which pertain to much larger regions in each galaxy)
are quite comparable to the total IR luminosities.

We can also compare the luminosity of the nuclear starburst to that
of a hidden AGN. We will do so using two techniques. First, we will
follow the procedure used by Heckman et al (1997) for Mrk 477
in which the low-ionization UV resonance emission-lines are attributed
to resonance-scattering of the UV continuum light from the hidden
Seyfert 1 nucleus off circum-nuclear gas clouds that cover a fraction $f$
of the sky as seen by the type 1 Seyfert nucleus.
The CII $\lambda$1335
emission-line in NGC 7130 and IC 3639 has a flux (and FWHM) of 
$1.6\times10^{-15}$ erg s$^{-1}$ cm$^{-2}$ (2.1 \AA) and
$1.4\times10^{-15}$ erg s$^{-1}$ cm$^{-2}$ (2.7 \AA) for NGC 7130 and IC
3639, respectively. Following Heckman et al (1997), we derive values for the
monochromatic UV continuum luminosity at 1335 \AA\
of the hidden Seyfert 1 nucleus of
2$\times10^8$ $f^{-1}$ L$\odot$ and 
$1.2\times10^8$ $f^{-1}$ L$\odot$ for NGC 7130 and IC 3639,
respectively, where f is the covering factor. The ratio of the relative numbers of
type 1 and type 2 Seyfert nuclei (in the context of the standard
unified model) implies that f $<$ 0.25.
We further adopt a bolometric correction factor of
5 (Elvis et al 1994), and therefore estimate that the lower limit to the
bolometric luminosity of the hidden Seyfert 1 nucleus is $3.9\times10^9$
L$\odot$ and $2.6\times10^9$ L$\odot$ for NGC 7130 and IC 3639,
respectively. Using this procedure, Heckman et al estimated a lower
limit to the luminosity of the hidden Seyfert 1 nucleus in Mrk 477
of $4\times10^{10}$ L$\odot$.
These values represent a lower limit to the luminosities
not only becuase of the upper limit to $f$,  but because
we have not applied any correction for intrinsic extinction
(we have applied only a correction for Galactic extinction). 

We can also use the case of NGC 1068, in which
the bolometric luminosity of the hidden Seyfert 1 nucleus
is about 10$^{11}$ L$\odot$, 
based on its reflected spectrum (Pier et al 1994).
We will refer these other four type 2 Seyfert galaxies to the
fiducial value for NGC 1068 using three tracers of the
intrinsic AGN luminosity: the [OIII]$\lambda$5007 emission-line,
the nuclear radio power at 1.4 GHz, and the nuclear mid-IR
monochromatic power at 10$\micron$ (Whittle 1992b; Heckman 1995).
Taking the log-average ratio of these three measures, we find
that the hidden Seyfert 1 nucleus in NGC 7130, NGC 5135, IC 3639,
and Mrk 477 has a bolometric luminosity that is 30\%, 20\%, 13\%,
and 150\% as luminous as NGC 1068 respectively. That is, we estimate
corresponding bolometric luminosities of $3\times10^{10}$ L$\odot$,
$2\times10^{10}$ L$\odot$, $1.3\times10^{10}$ L$\odot$, and
$1.5\times10^{11}$ L$\odot$. These are consistent with (several times larger
than) the lower limits to the luminosities from the CII $\lambda$1335 line
above.

Comparing these luminosities to those of the nuclear starburst (Table 5),
we see that
in these cases it appears that the bolometric luminosity of the
hidden type 1 Seyfert nucleus is comparable to that of the nuclear starburst 
It will be
important to see if this result of a rough scaling in the luminosity
of the type 1 Seyfert nucleus and nuclear starburst
is generally true in Seyferts.  

\section{Conclusions}

Our multiwavelength imaging and spectroscopic study of the 4 UV brightest Seyfert 2
 clearly shows that
nuclear (radii of $\sim10^{2}$ pc) starbursts are an important component in the
energetics of these nuclei.
They are the primary source of the so-called `featureless-continuum' 
in the UV-through-near-IR regime in these objects. In fact, this spectral component is
{\it not} featureless. We detect the direct spectroscopic signature of
massive stars: 1) stellar wind and photospheric lines in
our HST GHRS data, 2) HeI and high-order Balmer 
absorption-lines in the near-UV and visible and 3) CaII triplet absorption
in the near-IR due to red supergiants.

The estimated bolometric luminosity of the nuclear starburst
is roughly 10$^{10}$ L$_{\odot}$ in NGC 5135, IC 3639, and NGC 7130, and about
4$\times 10^{10}$ L$_{\odot}$ in Mrk 477.
The ratios between the starburst luminosity and the radio and [OIII]
luminosities in these galaxies are similar to those found in Mrk 477.
The difference found between the power of the starburst in Mrk 477 and 
that in NGC 7130, NGC 5135 and IC 3639 reflects the diference in one order of magnitud between nebular [OIII] emission line and the nonthermal radio continuum power of Mrk 477 with respect to the other three Seyfert 2 nuclei.
It seems that more powerful active nuclei are associated with more vigorous
nuclear starbursts (at least in this small sample). In fact, we
estimate that
the bolometric luminosity of nuclear starburst is similar to that of 
the AGN itself (the hidden type 1 Seyfert nucleus). In the case of Mrk 477,
the nuclear starburst plus the AGN may actually dominate the overall
energy output of the entire galaxy. In the other objects, the
sum of the nuclear starburst and AGN may contribute roughly 10 to 20\%
to the overall galaxy bolometric luminosity (which in all four cases is
primarily emitted in the mid/far-IR). The UV spectra taken through
the relatively large aperture of IUE 
imply that the most of the IR emission not due to the AGN plus
nuclear starburst is produced by a larger (several kpc-scale) dusty starburst.

The WFPC2 and FOC images suggest that the nuclear starburst has formed as
a consequence of the dynamics of the host galaxy. Bars (in NGC 7130, NGC
5135 and IC 3639) and/or interaction (Mrk 477 and NGC 7130) provide a very
efficient mechanism to drive the gas from the outer parts of the galaxy to
the nuclear region. Knots of star formation have been detected along the
leading edges of the bars and in spiral-like structures that extend into the
nucleus in these Seyferts 2 galaxies (as well as  in other low
luminosity active galaxies - Colina et al 1997). This flow of gas toward
the inner core may trigger star formation along the way, and probably
forms a
molecular torus that acts as a gas reservoir that obscures and
finally feeds the nucleus.
There, as has been suggested by Cid Fernandes \& Terlevich (1995) and
Heckman et al (1997), a nuclear starburst can naturally form from the
reservoir of molecular gas.

The four cases studied in this paper and in Heckman et al (1997) have been
selected on the basis of the brightness of the nucleus in the UV
(to enable spectroscopy with the GHRS).  This criterion may favor the selection of 
galaxies with more vigouros nuclear starbursts. Thus, an obvious and important question
is `Can this result be generalized to all the Seyfert 2 galaxies?' Our
preliminary analysis of ground-based near-UV and optical spectra of the
brightest 20 Seyfert 2 nuclei indicates that massive stars are definitely
present in the innermost arcsec of {\it at least} a-third of the nuclei.
These are just those cases where the strength of the young stellar continuum 
is greatest relative to the light from the old
underlying stellar population and from the nebular gas (e.g. in the cases
where the starburst continuum is most easily studied). Thus, we do not yet know whether nuclear starbursts
are present in all Seyfert nuclei or only in some. If the latter, is
there some key evolutionary connection between the starburst
and the Seyfert nucleus? The availibility of NICMOS and STIS on HST,
together with complementary multi-wavelength and ground-based data
may soon allow us to answer these fundamental questions.
 
% Authors may indicate to the editorial staff where they would like
% figures and tables to be placed in the manuscript.  This is done with
% either the \placefigure{KEY} or \placetable{KEY} commands.  These

% commands require \label{KEY} commands to be placed appropriately with
% corresponding table and figure captions.  When the manuscript is
% printed a short note is printed on the page where the figure or table
% is to go.  These commands are ignored in the aaspp4 and aas2pp4 styles.

%\placetable{tbl-3}
%\placefigure{fig1}

{\bf Acknowledgments}

We are grateful to Roberto Terlevich and Enrique P\'erez for many stimulating discussions and
helpful suggestions, and to the staff at STScI and at KPNO for their help 
in obtaining the data presented in this paper. This work was supported by
HST grants AR-05804.01-94A, GO-5944, and GO-6539 
from the Space Telescope Science Institute, which is operated by the
Association of Universities for Research in Astronomy, Inc., under NASA
contract NAS5-26555. It was also supported in part by the NASA LTSA
grant NAGW-3138.

\clearpage

\begin{deluxetable}{lccccccc}
\footnotesize
\tablecaption{Properties of the Seyfert 2 galaxies \tablenotemark{a}}
\tablewidth{0pt}
\tablehead{
\colhead{Name} & \colhead{Type} & \colhead{v} & \colhead{1 arcsec} & \colhead{log L$_{[OIII]}$} & 
\colhead{log L$_{1.4}$} & \colhead{L$_{FIR}$} & \colhead{L$_{IR}$} \nl
\colhead{} & \colhead{} & \colhead{km/s} & \colhead{pc} & \colhead{erg s$^{-1}$} & \colhead{erg s$^{-1}$} & \colhead{10$^{10}$ L$\odot$} & \colhead{10$^{11}$ L$\odot$}
} 
\startdata

 Mrk 477   & S       & 11340  & 730 & 42.74 & 39.34 & 5.5 & 1.5 \nl
 NGC 7130  & Sa pec  &  4842  & 310 & 41.30 & 39.00 & 10 & 1.8 \nl
 NGC 5135  & SBb     &  4112  & 270 & 41.05 & 38.95 & 7 & 1.6 \nl
 IC 3639   & SBbc    &  3285  & 210 & 40.76 & 38.46 & 2.3  & 0.5 \nl
\enddata
\tablenotetext{a} {Data of the [OIII] $\lambda$5007+4959 and 1.4 GHz radio continuum are from Whittle (1992b). FIR luminosity was computed from the 60 and 100 $\mu$m IRAS fluxes (Fullmer \& Lonsdale 1989); and IR luminosity computed using all 4 IRAS fluxes (Sanders \& Mirabel 1997). Radial velocities and morphological type are from NED. H$_0$=75 km s$^{-1}$ Mpc$^{-1}$}
\end{deluxetable}

\begin{deluxetable}{lcccccccc}
\footnotesize
\tablecaption{IUE flux at 2150 \AA\  and photometric properties from the FOC images
\tablenotemark{a}}
\tablewidth{0pt}
\tablehead{
\colhead{Name} & \colhead{F$_{IUE}$} & \colhead{F$_{FOC}$} & \colhead{A$_{2150}$} & 
\colhead{log P$_{tot}$} & \colhead{log P$_{nuc}$} & \colhead{log P$_{circum}$} & 
\colhead{$\mu_e$} & \colhead{R$_e$}  \nl
\colhead{1} & \colhead{2} & \colhead{3} & \colhead{4} & \colhead{5} & \colhead{6} & 
\colhead{7} & \colhead{8} & \colhead{9} 
} 
\startdata

 Mrk 477   & 4.2 & 2.1  & 0.27  & 43.20  & 41.77 & 42.89 & 14.33  & 160 \nl
 NGC 7130  & 6.8 & 3.9  & 0.45  & 42.80  & 41.61 & 42.62 & 13.95  &  80  \nl
 NGC 5135  & 8.9 & 6.2  & 0.92  & 43.05  & 41.55 & 42.88 & 15.18  & 200 \nl
 IC 3639   & 5.2 & 1.3  & 1.36  & 42.36  & 41.35 & 42.18 & 14.14  & 55 \nl
\enddata
\tablenotetext{a} {Column 2 and 3: Flux at 2150 \AA\ measured in the IUE spectrum (10 by 20 
arcsec aperture) and in the FOC images in units of 10$^{-15}$ erg s$^{-1}$ cm$^{-2}$ \AA$^{-1}$. 
The IUE fluxes are larger than the FOC fluxes implying that there is diffuse, low-surface-brightness 
emission that we are not detecting in the 14 by 14 arcsec FOC image (7 by 7 arcsec in the case of 
Mrk 477). 
Column 4: Galactic extinction at 2150 in magnitudes derived from the HI column density and 
N$_{HI}$/E(B-V) ratio.
Column 5, 6, and 7: The log of the  total, nuclear and circumnuclear monocromatic luminosity 
($\lambda$P$_{\lambda}$) at 2150 \AA\ in units of erg s$^{-1}$. These luminosities have been 
corrected only by the Galactic extinction. Column 8: The mean surface brightness in the STmag 
system interior to the effective half-light radii in units of mag arcsec $^{-2}$. 
Column 9: Half-light radii in pc.}
\end{deluxetable}

\begin{deluxetable}{lcccc}
\footnotesize
\tablecaption{Equivalent width (\AA) of the absorption features \tablenotemark{a}}
\tablewidth{0pt}
\tablehead{
\colhead{Line} & \colhead{type} & \colhead{NGC 7130} & \colhead{NGC 5135} & \colhead{IC 3639}  
} 
\startdata

 CIII $\lambda$1175    & S & 3.5 & 5.1 & --  \nl
 SiIII $\lambda$1205   & IS & 2.4 & --  & --  \nl
 NV $\lambda$1240      & W  & 6.1 & 4.6 & --  \nl
 SiII $\lambda$1260    & IS & 1.5 & 1.5 & 1.2   \nl
 SiII $\lambda$1263    & IS & 0.2 & 0.6 & 0.5  \nl
 OI+SiII $\lambda$1300 & IS & 4.0 & 4.5 & 1.95  \nl
 CII $\lambda$1335     & IS & 3.2 & 2.3 & 0.9   \nl
 SiIII+PIII $\lambda$1342& Ph & 1.0 & 0.6 & 0.4  \nl
 SiIV $\lambda$1400  & W+IS & 7.9 & 8.1 & 6.0 \nl
 SiIII $\lambda$1417   & Ph & 0.7 & 0.4 & --  \nl
 CIII $\lambda$1428    & Ph & 0.5 & 0.3 & 0.4  \nl
 SV $\lambda$1501      & Ph & 1.0 & 0.6 & 0.7 \nl
 SiII $\lambda$1526    & IS & 1.6 & 1.3 & 1.3 \nl
 CIV $\lambda$1550   & W+IS & 7.8 & 9.3 & 6.3 \nl
\enddata
\tablenotetext{a} {IS means Interstellar, W wind, and Ph Photospheric line.}
\end{deluxetable}

\begin{deluxetable}{lcccc}
\footnotesize
\tablecaption{Reddening estimates of the Seyfert 2 galaxies \tablenotemark{a}}
\tablewidth{0pt}
\tablehead{
\colhead{Name} & \colhead{E(B-V)$_{Gal}$}  & \colhead{E(B-V)$_{UV}$}  & \colhead{E(B-V)$_{IUE}$} &
 \colhead{E(B-V)$_{Bal}$}  
} 
\startdata

 Mrk 477   & 0.028  & 0.16 & 0.30 & 0.23 \nl
 NGC 7130  & 0.046  & 0.25 & 0.56 & 0.75 \nl
 NGC 5135  & 0.094  & 0.20 & 0.53 & 0.67 \nl
 IC 3639   & 0.138  & 0.20 & 0.22 & 0.52 \nl
\enddata
\tablenotetext{a} {Column 2: Foreground extinction from the Milky Way. Uses the HI column density
 from Stark et al (1992) and assumes N$_{HI}$/E(B-V)=4.93x10$^{21}$ cm$^{-2}$.
Column 3: Intrinsic extinction for the GHRS spectrum estimated from comparing the spectral energy
 distribution of this spectrum to the synthetic starburst spectrum that reproduces the stellar 
absorption features. Assumes the Calzetti et al (1994) UV effective extinction law. 
Column 4: Intrinsic extinction for IUE spectrum, estimates from comparing the spectral energy 
distribution of this spectrum to a generic synthetic starburst with a slope $\beta$=-2.5. 
Column 5: Intrinsic extinction derived from the emission-line gas. See text for details.}
\end{deluxetable}

\begin{deluxetable}{lcccccccc}
\footnotesize
\tablecaption{Intrinsic properties of the starbursts \tablenotemark{a}}
\tablewidth{0pt}
\tablehead{
\colhead{Name} & \colhead{log L$_{1500}$} & \colhead{Mass}  & \colhead{L$_{Bol}$(SB)} & 
\colhead{L$_{Bol}$(IUE)} & \colhead{L$_{Bol}$(Nuc)} & \colhead{log Q} & \colhead{log Q} &
 \colhead{log Q} \nl
\colhead{1} &  \colhead{2} & \colhead{3} & \colhead{4} & \colhead{5} & \colhead{6} & \colhead{7}
 & \colhead{8} & \colhead{9} 
} 
\startdata

 Mrk 477   & 40.75  & 49  & 3.6 & 24  & 4.0-15  & 53.6 & 54.0  &  --\nl
 NGC 7130  & 40.26  & 9   & 1.1 & 30  & 0.4-3   & 53.4 & 53.4  & 53.7 \nl
 NGC 5135  & 40.24  & 9   & 1.1 & 33  &    - 2  & 53.3 & 53.4  & 53.7  \nl
 IC 3639     & 39.74  & 3   &  0.3 & 3.6 & 0.3-1.3 & 52.8 & 52.9  & -- \nl
\enddata
\tablenotetext{a} {Column 2: The log of the monocromatic luminosity at 1500 \AA\ in the GHRS aperture in units of erg s$^{-1}$ \AA$^{-1}$. This assumes the corrections for Galactic and intrinsic extinction given in Table 4.
 Column 3: The implied mass (units of 10$^6$ M$\odot$) of the starburst inside the GHRS aperture assuming an instantaneous burst of star formation with an age of 4 Myr (except for Mrk 477, which is 6 Myr), solar metallicity, a Salpeter IMF with an upper mass cut-off of 100 M$\odot$ and a lower mass cut-off of 1 M$\odot$. 
Column 4: The implied bolometric lumninosity of the starburst inside the GHRS aperture in units of 10$^{10}$ L$\odot$ . 
Column 5: As in column 4 inside the IUE aperture in units of 10$^{10}$ L$\odot$. 
Column 6: The bolometric luminosity of the nucleus in units of 10$^{10} $ L$\odot$.  The two values are derived using two different techniques. See text for details.
Column 7: The log of the Lyman ionizing photon luminosity (units of phot s$^{-1}$) inside the GHRS aperture based on the extinction-corrected UV luminosity in Col. 2 and assuming the same burst models than in Col. 3. 
Column 8: As in column 7 based on the observed Balmer emission-lines and the extinction correction given in Col. 5 of Table 4. 
Column 9: As in Col. 7 based on the observed Br$\gamma$ emission (Goldader et al 1997). }
\end{deluxetable}

% And finally, we must deal with the figures.  There are three figures
% associated with this manuscript; two figures are Encapsulated
% PostScript (EPS) files.  The third figure is a grey scale figure that does
% not exist in EPS form.
%
% Authors have three options for including figure information within a
% manuscript.  Not all the options may be acceptable by the target Journal - be
% sure to look at the appropriate submission instructions, electronicor
% otherwise.
%
% Option 1.  Using this option, only the figure captions are included in the
% main body of the manuscript.  The figure captions must start on a new page.
% The captions are generated with the \figcaption[]{} command: the first
% argument is optional, if you put something in there, put the name of the
% EPS file that goes with the caption; the second argument is the figure
% caption itself, and may include a \label command.  The \figcaption command
% generates the figure numbers.  This option is acceptable for all manuscript
% submissions.

\clearpage

%Figure 1
\figcaption{The WFPC2 F606W images. The orientation in all the images is North
 up and East to the left. The images show the morphological characteristics of
the inner galaxy: a) NGC 7130 shows dust lanes running from North to South. 
The displayed field size is roughly 9 by 9 kpc. b) NGC 5135 shows arc-like
segments spiraling to the center of the galaxy. The displayed field size is
roughly 6.5 by 6.5 kpc.
c) IC 3639 shows its bar and its spiral arms emerging from the edge of the bar.
The displayed field size is roughly 6.7 by 6.3 kpc}

%Figure 2
\figcaption{The UV surface brightness profiles of the new galaxies in the
sample.  These were measured from the FOC $512z \times 1024$ images, after
sky subtraction, and using circular annuli centered on the origins mentioned
in the text.  The solid line shows the mean surface brightness within the
annuli, while the dotted lines show the effects of changing the adopted sky
level by its $\pm 1\sigma$ uncertainty. The arrow indicates the adopted outer
radius of the inner high surface brightness structure.}

%Figure 3
\figcaption{The FOC F210M image of the ultraviolet continuum in NGC 7130. The field displayed (roughly 2.6 by 2.9 kpc)
shows knots that are located in the 
arc-like segments that could be associated with the leading edge of the bar.}

%Figure 4
\figcaption{The central 2$\times$2 arcsec$^2$ (620 by 620 pc) of NGC 7130 in optical 
light (a) and in UV light (b). The origin of the plot is the knot with the
highest UV surface brightness. 
The presumed nucleus is placed at 0.2 arcsec West and 0.1 arcsec South of the
 origin.}

%Figure 5
\figcaption{Overall view of the UV continuum emission detected in the FOC image of NGC 5135. The arc structure is coincident with the arc segment seen in the optical WFPC2 image that is spiraling to the center. The displayed field size is
about 2.2 kpc.}

%Figure 6
\figcaption{Central 3$\times$3 arcsec$^2$ (540 by 540 pc)
field of NGC 5135: a) WFPC2, b) UV. The origin of this plot is the pixel with
the highest UV surface brigtness. Based on the morphology of the WFPC2 image,
we believe that the nucleus is the knot located at 1.05 arcsec North and 0.4 arcsec East.}

%Figure 7
\figcaption{Central 1$\times$1 arcsec$^2$ (210 by 210 pc) field 
of IC 3639: a) WFPC2, b) UV. The origin of this plot is the center of the ${\bf S}$ structure seen in the WFPC2 image, which is coincident with the pixel of the highest surface brightness in the central 0.1$\times$0.1
arcsec of the UV image. We suggest that the nucleus of the galaxy is placed there.}

%Figure 8
\figcaption{The GHRS spectra of: a) NGC 7130, b) NGC 5135, c) IC 3639. They are
displayed in log (F$_{\lambda}$) to show the emission and absorption lines. The
most important stellar wind and photospheric absorption lines in the galaxy 
(NGC 7130) are labeled.  }

\figcaption{The normalized GHRS spectrum of IC 3639 and one simulation in which scattered light from a hidden Seyfert 1 nucleus produces 5$\%$ of the UV continuum. (dashed line)}

\figcaption{$\chi^2$ parameter of the fits to the profile of SiIV in 
NGC 7130 for burst models (a,b) and continuous star formation models (c,d) with different M$_{upp}$ (a,c) and IMF slope (b,d). 
The size of the bubble is proportional to $\chi^2$, with a larger size indicating a worse fit. The scale in the small boxes indicates the value of $\chi^2$.}

\figcaption{As in Figure 12 for NGC 5135}

\figcaption{As in Figure 12 for  IC 3639}

\figcaption{The GHRS spectrum of NGC 7130 dereddened by E(B-V)=0.046 using the MW extinction law and by E(B-V)=0.25 using the Calzetti et al (1994) extinction law (thick line) and the synthetic 4 Myr burst model (in relative units) (thin line). The IMF slope is Salpeter and M$_{upp}
$=80 M$\odot$.}

\figcaption{The GHRS spectrum of NGC 5135 dereddened by E(B-V)=0.094 using the MW extinction law and by E(B-V)=0.2 using the Calzetti et al (1994) extinction law (thick line) and the synthetic 5 Myr burst model (in relative units) (thin line). The IMF slope is Salpeter and M$_{upp}$
=80 M$\odot$.}

\figcaption{The GHRS spectrum of IC 3639 dereddened by E(B-V)=0.138 using the MW extinction law and E(B-V)=0.2 using the Calzetti et al (1994) extinction law (thick line) and the synthetic 4 Myr burst model (in relative units) (thin line). The IMF slope is Salpeter and M$_{upp}$=60 M$\odot$.}

\figcaption{Ground-based optical spectrum of NGC 7130 (a), NGC 5135 (b) and IC 3639 (c) obtained through a 1.5$\times$3.5 arcsec$^2$ aperture displayed as log (F$_{\lambda}$) to show the emission and absorption lines. The inset shows the region around 3650-3850
 \AA\ expanded to highlight the high-order Balmer series and HeI $\lambda$3819
lines. Other photospheric absorption lines are labeled in the figure. The wavelength axis is in the rest frame of the galaxy.}

\figcaption{The normalized spectrum of NGC 7130 (dashed line) plotted with the normalized spectrum of a B0V star combined with a G0V star (thick line). 60$\%$ of the light is from a B0V and 40$\%$ from a G0V star. The comparison shows that most of the
stellar features in NGC 7130 are well reproduced by a combination of young (B0V) and old (G0V) stars.}

\figcaption{The normalized spectrum of NGC 5135 (dashed line) plotted with the normalized spectrum of the super-star cluster B in NGC 1569 (thick line). 
The stellar features in NGC 1569 B are well
reproduced by a combination of young (70 $\%$ of the light from B0V) and old (30 $\%$ of the light from G0V) stars. There is a good match between NGC 5135 and NGC 1569 B.}

\figcaption{The normalized spectrum of IC 3639 (dashed line) plotted with the combination of the normalized spectrum of the super-star cluster B in NGC 1569 and the galaxy NGC 4339 (thick line). 60$\%$ of the light is from NGC 1569 and 40$\%$ from
NGC 4339. The stellar lines of NGC 4339 are the typical features from an old bulge population.}

\figcaption{The CaII triplet region in NGC 7130. The absorption features and some emission lines are labeled.}

\figcaption{Profile of the emission lines in NGC 5135: a) H$\beta$, b) [OIII], c) HeII $\lambda$4686. The observed profile (solid line), the two model components (dashed lines) and their sum (solid line) are shown.}

\end{document}